\documentclass[trackchanges,twocolumn]{aastex7}

\usepackage{booktabs}
\usepackage{caption}
\usepackage{multirow}
\usepackage{amsmath}
\usepackage{subcaption}
\usepackage{units}
\begin{document}

\title{A Census of Pulsars in Possible Association with Galactic Open Clusters}

\author[0009-0000-1929-7121]{Lu Zhou}
\affiliation{School of Physics and Technology, Wuhan University, Wuhan, Hubei 430072, China}
\affiliation{Department of Physics, Faculty of Arts and Sciences, Beijing Normal University, Zhuhai 519087, China}
\affiliation{Advanced Institute of Natural Sciences, Beijing Normal University, Zhuhai 519087, China}
\email{zhoulugz@whu.edu.cn}

\author[0000-0002-3309-415X]{Zhi-Qiang You}
\affiliation{Institute for Gravitational Wave Astronomy, Henan Academy of Sciences, Zhengzhou 450046, Henan, China}
\email{zhiqiang.you@hnas.ac.cn}

\author[0000-0002-0880-3380]{Lu Li}
\affiliation{Shanghai Astronomical Observatory, Chinese Academy of Sciences, 80 Nandan Road, Shanghai 200030, China}
\email{lilu@shao.ac.cn}

\author[0000-0002-2187-4087]{Xiao-Jin Liu}
\affiliation{Department of Physics, Faculty of Arts and Sciences, Beijing Normal University, Zhuhai 519087, China}
\affiliation{Advanced Institute of Natural Sciences, Beijing Normal University, Zhuhai 519087, China}
\email[show]{xliu@bnu.edu.cn}

\author[0000-0001-7049-6468]{Xing-Jiang Zhu}
\affiliation{Department of Physics, Faculty of Arts and Sciences, Beijing Normal University, Zhuhai 519087, China}
\affiliation{Advanced Institute of Natural Sciences, Beijing Normal University, Zhuhai 519087, China}
\affiliation{Institute for Frontier in Astronomy and Astrophysics, Beijing Normal University, Beijing 102206, China}
\email[show]{zhuxj@bnu.edu.cn}

\author[0000-0002-3567-6743]{Zong-Hong Zhu}
\affiliation{School of Physics and Technology, Wuhan University, Wuhan, Hubei 430072, China}
\email[show]{zhuzh@whu.edu.cn}


\begin{abstract}
Among the $\sim 4000$ known pulsars in our Galaxy, $\lesssim 10\%$ are found in globular clusters, but none has been confirmed in any open clusters yet, although they outnumber globular clusters by about 20 times.
In this work, we make use of the Gaia DR3 catalog of Galactic open clusters and conduct a pulsar census, in order to identify pulsars that are either 1) current members of open clusters, or 2) escaped from open clusters to the field.
Among 164 pulsars with independent distance measurements and 3530 open clusters, we find that 4 pulsars are likely residing in open clusters. In particular, we find compelling evidence that the binary pulsar J1302$-$6350 (B1259$-$63) is a member of the open cluster UBC~525; based on Gaia data, we update its distance to be $2.26\pm 0.07$~kpc and measure the mass of its companion Be star LS 2883 to be $\unit[16.8]{ M_\odot}$.
For 145 pulsars with both distance and proper motion measurements and 2967 open clusters with full kinematic parameters, we trace the past trajectories of both pulsars and open clusters in the Galactic gravitational potential, and find pulsars that were within 3 times the radius of a cluster.
This results in 19 pulsars that were likely born in open clusters.
We discuss implications for the formation history of PSR J1302$-$6350 and highlight the scientific potential of searching for pulsars in open clusters.

\end{abstract}


\section{Introduction} \label{sec:introduction} 

Pulsars are rapidly rotating neutron stars formed through the core collapse of massive stars at the end of their life cycles.
They are fascinating astronomical objects that serve as unparalleled natural laboratories for testing fundamental physics under extreme conditions.
The formation of pulsars, or neutron stars in general, in supernova explosions also offers a unique window into the endpoints of stellar evolution.
Currently more than 4000 pulsars have been discovered in our Galaxy; see the latest version of the ATNF Pulsar Catalogue\footnote{\url{https://www.atnf.csiro.au/research/pulsar/psrcat}} \citep{MHT+05}.
Among them, pulsars found in Galactic globular clusters account for nearly 10\% of the total population; the fraction is high given that globular clusters make up $\ll 1\%$ of the Galactic stellar mass.

However, no pulsars have been definitively found in any open clusters in the Milky Way.
There are a couple of factors that make open clusters unfavorable places to find a pulsar.
First, open clusters are relatively low-mass and loosely bound gravitational systems that are susceptible to tidal disruption and typically dissolve on timescales of a few hundred million years.
Their lifespans could be shorter than the time required for a massive star to evolve, undergo a supernova, and leave behind a neutron star that would be observable as a pulsar.
Second, even if a pulsar were to form, the violent ``kick" imparted by the asymmetric supernova explosion might eject the pulsar into the Galactic field.

Recently, \citet{LSB+25} performed the first search for pulsars in seven old massive open clusters using the Five-hundred-meter
Aperture Spherical radio Telescope (FAST), and found PSR~J1922+37 in the direction of NGC 6791; the pulsar distance informed by Galactic electron density model is consistent with the cluster distance, suggesting a possible physical association.
This motivated \citet{ZhangSB25} to search for new pulsars in four old open clusters using archival observations taken with the Parkes radio telescope; they found no new pulsars but detected three rotating radio transient candidates.

As gravitationally bound collections of stars born from the same giant molecular cloud, open clusters offer a snapshot of stellar evolution with a well-defined age, chemical composition, and dynamical history. The discovery of a pulsar within such a system would provide an unparalleled opportunity to anchor its properties to a known stellar population. This association allows for a direct comparison between the pulsar's characteristic spin-down age and the cluster's robust isochronal age, constraining models of pulsar evolution. Furthermore, it could enable the estimation of the progenitor star's mass from the cluster's initial mass function and offers a chance to study the impact of a supernova on the cluster's gas and stellar dynamics.

In this companion paper to \citet{LSB+25}, we cross match the known pulsar population and the open cluster catalog. The goal is to identify pulsars that can be associated with open clusters either at present or in the past.
The transformative astrometric data from the Gaia mission have made such a study feasible. Gaia Data Release 3 (DR3) provided precise parallaxes and proper motions for over a billion stars, leading to the discovery and precise characterization of thousands of new open clusters and their members. This, combined with an expanded catalog of pulsars with measured proper motions and independently constrained distances, allows us to conduct a systematic, Galaxy-wide census.
Through this analysis, we aim to establish a catalog of candidates, quantify the likelihood of these associations, and shed new light on the birth environments and kick velocities of neutron stars.

This paper is organized as follows. 
In \autoref{sec:Data}, we describe the catalogs of pulsars and open clusters used in this work.
In \autoref{sec:Method}, we describe the calculation details. We first search for present-day spatial and kinematic coincidences among pulsars and open clusters while accounting for measurement uncertainties.
We then perform kinematic back-integration of pulsar and cluster orbits in a realistic Galactic potential to uncover past ejection events.
In \autoref{sec:Result}, we present the candidate pulsar-open cluster associations found in our analysis. 
In \autoref{sec:Discussions}, we discuss the top candidates in more detail and their implications. 
Finally, we present concluding remarks in \autoref{sec:Conclusions}.


\section{Catalogs of pulsars and open clusters} \label{sec:Data}

\subsection{Pulsar Data}\label{sec:psr-data}

Distance is a crucial parameter in distinguishing accidental overlap from genuine associations between a pulsar and an open cluster.
For the present-day association calculations, we only consider
pulsars with independent distance measurements. These data are compiled from the following three sources: 
\begin{enumerate}
    \item The collection of 162 pulsar distances listed in \cite{YMW17}, excluding those in  globular clusters or the Magellenic clouds.
    \item 77 millisecond pulsars with parallax distances in the MeerKAT pulsar timing array \citep{SBF+24}, of which 35 have statistical significance greater than $3\sigma$.
    \item The Cornell catalog\footnote{\url{https://hosting.astro.cornell.edu/research/parallax}} of pulsar parallax obtained with very long baseline interferometry (VLBI).
\end{enumerate}
We note that 74 pulsars are present in more than one reference and have multiple distance measurements. 
For these pulsars, we only use the distance with the highest precision. 
In total, we obtained 248 pulsars with reliable distance constraints. After further selection based on a statistical significance greater than $3\sigma$, we obtain 164 pulsars, which are listed in \autoref{tab:pulsar_independent_dist} of \autoref{appendix_distance}. 
Other pulsar parameters, including position, spin period and age, are obtained from the ATNF Pulsar Catalogue. 

For the past-trajectory calculations, complete astrometric parameters (position, distance, age, proper motion, and radial velocity) are essential. Among the 164 pulsars with distance measurements, we exclude those without proper motions by querying the ATNF Pulsar Catalogue (version 2.4.0), leading to 145 pulsars for our pulsar birth-place investigations. 
As radial velocities are typically unknown for pulsars, we adopt two velocity distributions.
Our default model is a uniform distribution between $-200$ and $200 \, \mathrm{km \, s^{-1}}$, and the alternative model is from \cite{Hobbs2005}: $152 \pm 10 \, \mathrm{km \, s^{-1}}$ for normal pulsars and $54 \pm 6 \, \mathrm{km \, s^{-1}}$ for recycled pulsars.
We also test how sensitive our results are to the adopted radial velocity distribution.

\subsection{Open cluster Data}\label{sec:oc-data}

\cite{HR24} analyzed the most recent data from Gaia DR3 and compiled a comprehensive catalog of likely Galactic star clusters, which include both globular and open clusters, as well as moving groups. 
Moving groups may originate from open clusters and are characterized as a set of stars with similar kinematic properties, ages, and chemical compositions. However, due to tidal forces or other dynamical effects, they gradually disperse and become unbound, no longer a tightly concentrated group \citep{JHB96, HR24}. 
Given their kinematic similarities, we consider 3530 high-quality open clusters and 539 moving groups from \cite{HR24} in our analysis. 
We adopt the key cluster parameters from \cite{HR24}, including position, distance, proper motion, characteristic size, and age, in the calculations that aim to identify pulsars currently associated with open clusters or moving groups. 
For pulsar birth-place investigations, we only consider open clusters in the Gaia DR3 catalog; after leaving out those without radial velocity measurements, we have 2967 clusters with complete kinematic parameters (position, distance, proper motion, and radial velocity) and physical properties (radius and age).

\section{Method} \label{sec:Method}

\subsection{Searching for pulsars associated with open clusters at present}\label{sec:method-current}

To identify associations between pulsars and open clusters at the current epoch, our method can be divided into the following three steps. 

First, we calculate the angular distance ($\theta$) between each pulsar and each open cluster and select pairs when $\theta < 3r_{50}$, where $r_{50}$ is the angular radius that covers 50\% of the members of the cluster. We chose $3r_{50}$ as the selection criterion to ensure the inclusion of a highly complete population of cluster members. 
According to \cite{LS22},  for a typical open cluster (assuming a \cite{K62} profile with $r_t/r_c\simeq2$), a selection radius of $3r_{50}$ is expected to include 99\% of the member stars. For an extremely loose open cluster ($r_t/r_c\simeq20$), it will include 90\% of the member stars. Here $r_t$ and $r_c$ are the tidal radius and core radius, respectively. 

Then, for the selected pairs, we compare the measured distances of the pulsars with those of the open clusters. 
The pairs are saved if the distance measurements of both targets are consistent within $5\sigma$, giving candidate pairs that are spatially overlapped. 
For these candidate pairs, we estimate the association probability using a Monte Carlo approach. 
Since the positions (RA, Dec) of both pulsars and open clusters are precisely measured, we ignore their positional uncertainties. 
To account for uncertainties of distance measurements, we assume the distances of both pulsars and open clusters follow Gaussian distributions, with the means and standard deviations for pulsars listed in \autoref{tab:pulsar_independent_dist}, and the cluster distances adopted from \cite{HR24}. 
We generate $10^7$ random distance samples for each pulsar and open cluster, and calculate the linear separation between them. 
If the separation was less than the linear projection of $3r_{50}$, it is counted as an association. The association probability is then computed as $\text{Prob} = N_{\text{3r50}} / 10^7$, where $N_{\text{3r50}}$ is the number of such occurrences. 

Finally, we compare the flight distance of the pulsar ($s$) with the characteristic size of the open cluster, that is, $s =\Delta \mu \tau_{\rm psr} \le 3r_{50}$, where $\Delta \mu$ is the difference in proper motion and $\tau_{\rm psr}$ is the characteristic age of the pulsar. 
Pairs survived all the tests above are then saved for further analysis.

All computations are performed using the {\tt\string SkyCoord} module in the \textsc{astropy} software package \citep{Astropy13, Astropy22}, which adopts a default Galactocentric Solar distance of 8.122 kpc \citep{DP18} and a Solar height of 20.8~pc above the Galactic plane \citep{2019_solar}.

\begin{figure}[t]
    \centering
    \includegraphics[width=0.46 \textwidth, trim=0 0 0 0, clip]{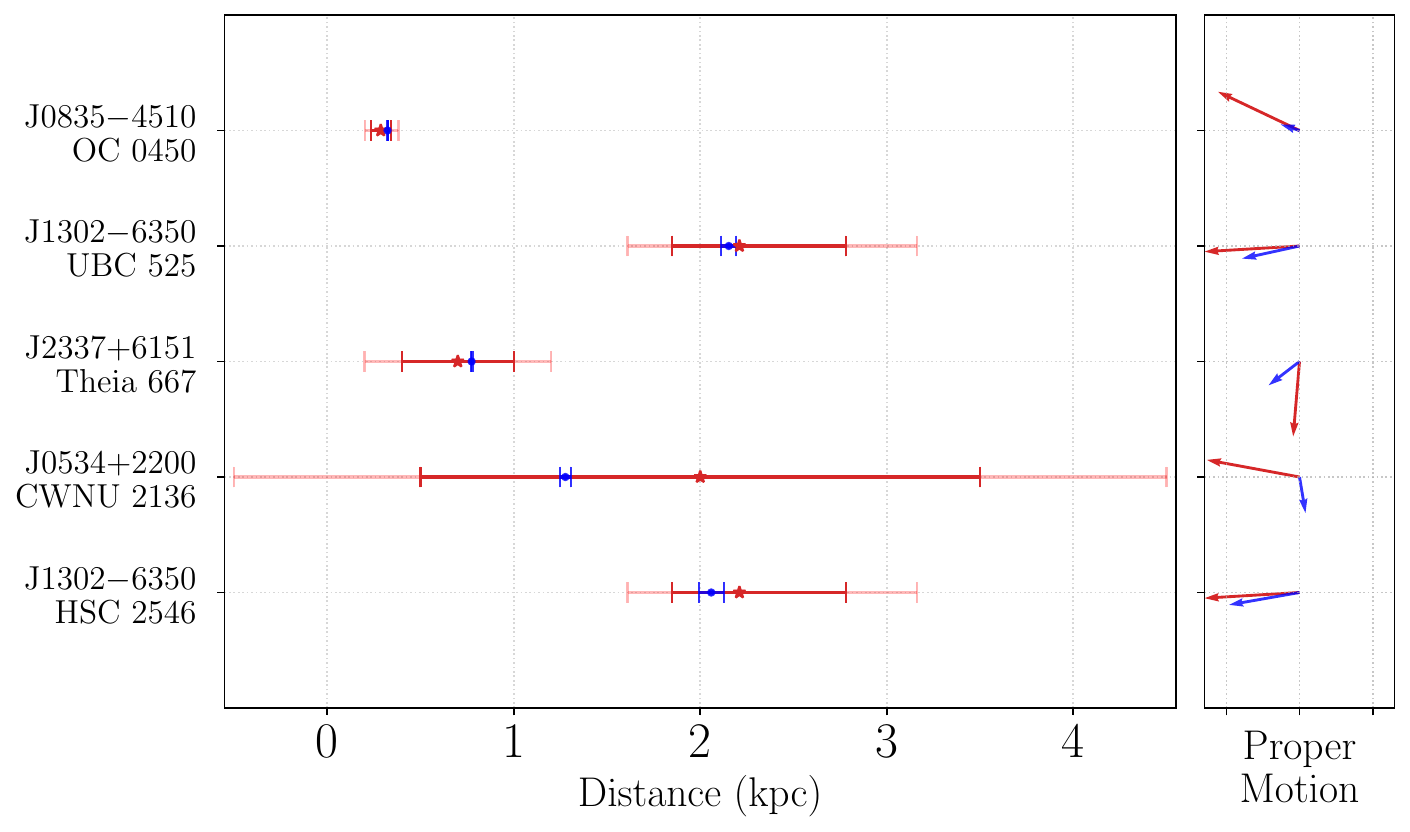}
    \caption{Distance and proper motion distributions for candidate pulsar-open cluster associations. 
    Left panel: Each row represents a pulsar-cluster pair, with open cluster distances shown as blue error bars ($\pm5 \sigma_{\text{oc}}$, where $\sigma_{\text{oc}}$ is the cluster distance uncertainty) and pulsar distances marked by red ($\pm3 \sigma_{\text{psr}}$) and light-red ($\pm5 \sigma_{\text{psr}}$) error bars with $\sigma_{\text{psr}}$ being the pulsar distance uncertainty.
    Right panel: Proper motion vectors for open clusters (blue arrows) and pulsars (red arrows). Each row corresponds to the pair in the left panel, while segment lengths indicate proper motion magnitudes, and arrows show directions.}
    \label{fig:psr_oc_dist_pm}
\end{figure}

\begin{table*}[t]
    \centering
    \setlength{\tabcolsep}{6pt} 
    \renewcommand{\arraystretch}{1.1} 
    \begin{tabular}{ccccccccc}
        \hline\hline
        Name & Prob & $d$ & $\mu_{\alpha} \cos \delta$ & $\mu_{\delta}$ & Age & $P_0$ & $r_{50}$ & $M_{\rm tot}$ \\
         & ($\%$) & (kpc) & ($\operatorname{mas~yr}^{-1}$) & ($\operatorname{mas~yr}^{-1}$) & (yr) & (s) & (deg/pc) & ($\mathrm{M}_{\odot}$) \\
        \midrule 
        J0835$-$4510 & \multirow{2}{*}{36.4} & $0.287^{+0.019}_{-0.017}$ & $-49.68 \pm 0.06$ & $29.90 \pm 0.10$ & $1.13 \times 10^{4}$ & 0.09 & -- & --\\ 
        OC 0450 & & $0.3236 \pm 0.0001$ & $-11.40 \pm 0.02$ & $4.66 \pm 0.02$ & $2.17 \times 10^{7}$ & -- & 3.14/17.18 & 251.32 \\ 
        \midrule
        J1302$-$6350 & -- & $2.21^{+0.19}_{-0.12}$ & $-7.01 \pm 0.03$ & $-0.53 \pm 0.03$ & $3.32 \times 10^{5}$ & 0.05 & -- & --\\ 
        UBC 525 & 10.3 & $2.1530 \pm 0.0082$ & $-4.24 \pm 0.01$ & $-1.18 \pm 0.01$ & $9.03 \times 10^{6}$ & -- & 0.16/6.17 & 1457.06 \\
        HSC 2546 & 1.4 & $2.0590 \pm 0.0136$ & $-5.20 \pm 0.02$ & $-1.16 \pm 0.01$ & $2.45 \times 10^{7}$ & -- & 0.23/8.22 & 534.70 \\
        \midrule 
        J2337+6151 & \multirow{2}{*}{8.1} & $0.7 \pm 0.1$ & $-1 \pm 18$ & $-15 \pm 16$ & $4.06 \times 10^{4}$ & 0.50 & -- & -- \\ 
        Theia 667 & & $0.7742 \pm 0.0009$ & $-4.88 \pm 0.02$ & $-4.76 \pm 0.03$ & $1.24 \times 10^{8}$ & -- & 0.83/11.23 & 395.89 \\ 
        \midrule 
        J0534+2200 & \multirow{2}{*}{6.0} & $2.0 \pm 0.5$ & $-11.34 \pm 0.06$ & $2.65 \pm 0.14$ & $1.26 \times 10^{3}$ & 0.03 & -- & --\\ 
        CWNU 2136 & & $1.2767 \pm 0.0057$ & $0.73 \pm 0.02$ & $-5.57 \pm 0.02$ & $3.31 \times 10^{8}$ & -- & 1.99/44.40 & 375.45 \\ 
        \midrule 
    \end{tabular}
    \caption{Parameters of pulsars and their possibly associated open clusters. Name: Pulsar name (above) and cluster name (below). Prob: Association probability. $d$: Distance. $\mu_{\alpha} \cos \delta$, $\mu_{\delta}$: Proper motion components in right ascension and declination. Age: Pulsar spin-down age, and open cluster age. $P_0$: Pulsar spin period. $r_{50}$: Radius containing 50\% of the cluster members. $M_{\rm tot}$: Cluster total mass.}
    \label{tab:current_psr_oc_pair}
\end{table*}

\subsection{Construction of past trajectories of pulsars and open clusters}\label{sec:method-past} 

We compute the past trajectories of pulsars and open clusters through numerical orbit integration using {\tt\string Gala}\footnote{\url{https://gala.adrian.pw/en/latest/}} \citep{P17} with the {\tt\string MWPotential2014} Galactic potential \citep{B15}. 
To determine the true Galactic motions of pulsars and open clusters, the Sun's peculiar motion must be subtracted from their observed velocities. The Solar velocity with respect to the Local Standard of Rest (LSR) is given by: $\mathbf{V}_{\odot}$ = (12.9, 245.6, 7.78)~km~s$^{-1}$ \citep{DP18}, while the Sun’s distance to the Galactic center and height above the Galactic plane are the same as in Section~\ref{sec:method-current}. 

To account for observational uncertainties, we conduct Monte Carlo simulations by sampling distance, proper motion, and radial velocity for both the pulsar and cluster within their measurement errors.
We first generate $10^4$ samples in the parameter space.
In each trial, we trace the backward trajectories of the pulsar and the cluster using a time step of $0.002 {T_{\text{end}}}$, where ${T_{\text{end}}}$ is the smaller of the pulsar's age and the cluster's age. 
During the orbital integration, we evaluate the pulsar-cluster separation $\Delta$ at each time step, defining a close encounter as when the minimum separation $\Delta_{\text{min}}$ falls below 3 times the cluster's radius ($\Delta_{\text{min}} < 3r_{50}$). 
The encounter times $\tau$ corresponding to these minimum separations are recorded for analysis. 

The resulting $\Delta_{\text{min}}$ distribution follows the absolute difference of two 3D Gaussian distributions, with expectation values and standard deviations $(m_{1}, s_{1})$ and $(m_{2},s_{2})$, as described by \cite{HBZ01, THS+12}: 
\begin{multline}\label{eq:1}
W_{3 \mathrm{D}}(\Delta_{\text{min}}) = \frac{\Delta_{\text{min}}}{\sqrt{2\pi} s m}\times \\
\left\{\exp \left[-\frac{1}{2} \frac{(\Delta_{\text{min}}-m)^{2}}{s^{2}}\right]
- \exp \left[-\frac{1}{2} \frac{(\Delta_{\text{min}}+m)^{2}}{s^{2}}\right]\right\},
\end{multline} 
where $m=\left|m_{1}-m_{2}\right|$ and $s^{2}=s_{1}^{2}+s_{2}^{2}$. 

If pulsar and open cluster are at the same place in the past (i.e., $m \to 0$), Equation~\ref{eq:1} simplifies to:

\begin{equation}\label{eq:2}
W_{3 \mathrm{D}, m \to 0}(\Delta_{\text{min}})=\frac{2 \Delta_{\text{min}}^{2}}{\sqrt{2 \pi} s^{3}} \exp \left(-\frac{\Delta_{\text{min}}^{2}}{2 s^{2}}\right).
\end{equation}

Even in the ideal case, the probability of finding $\Delta_{\text{min}}=0$ in Monte Carlo simulations is zero. However, for a large number of simulations, values close to zero may be found. For pulsar-cluster pairs with $\Delta_{\text{min}}$ less than three times the cluster radius for at least 1 out of $10^4$ Monte Carlo trials, we increase the number of samples to $10^6$ to calculate the probability of close encounter.

\section{Results} \label{sec:Result}
\subsection{Possible Pulsar-Open Cluster Associations at Present}
\label{sec:results-present}

Cross matching 164 pulsars with 3530 open clusters, we identified 5 possible pulsar-cluster pairs with present-day association probability higher than $1\%$.
Below we comment on the interpretation of the probability and our chosen threshold.
First, the calculated probability is simply the probability of spatial overlap between a pulsar and an open cluster, and it does not necessarily reflect the probability for a physical association. To establish highly probable pulsar-open cluster associations would require more in-depth investigations, for which we illustrate in the following section.
Second, the apparently low threshold of $1\%$ allows candidates reported in this study to be revisited once higher precision astrometry data become available.
Third, by performing the same spatial-overlap calculations for member stars of Gaia open clusters, we find that a probability of a few percent is typical for some clusters (e.g., UBC 525 and HSC 2546). This means that our threshold is not too permissive.

In \autoref{fig:psr_oc_dist_pm}, we show the distance distributions and proper motion characteristics of 5 pulsar-open cluster pairs, in an order of the association probability.
\autoref{tab:current_psr_oc_pair} summarizes their key parameters.
As shown in \autoref{fig:psr_oc_dist_pm}, the distance measurements of these pulsar-open cluster pairs are consistent within $2\sigma_{\text{psr}}$. PSR J0835$-$4501 and OC~0450 have the highest association probability ($36\%$), due to the large radius of OC~0450 and the small spatial separation between the pulsar and the cluster.
For PSR J1302$-$6350, both UBC~525 and HSC~2546 lie within the $1\sigma_{\text{psr}}$ range of the pulsar distance, with association probabilities of $10\%$ and $1\%$, respectively. Notably, UBC~525 is spatially closer to J1302$-$6350 compared to HSC~2546, and its mass ($\approx1457$~$\mathrm{M}_{\odot}$, \citealp{HR24}) makes it the most massive cluster among the five studied.

In the right panel of \autoref{fig:psr_oc_dist_pm}, the proper motions of J1302$-$6350 \& UBC~525 and J1302$-$6350 \& HSC~2546 exhibit closer agreement in both magnitude and direction compared to the other pairs.
To further examine these associations, we plot in \autoref{fig:current_psr_oc_coord} of \autoref{append:psr-oc} the 3D spatial distributions of top three candidate pulsar-cluster pairs listed from \autoref{tab:current_psr_oc_pair}.
From \autoref{fig:current_psr_oc_coord}, one can see that J1302$-$6350 is positioned near the center of its associated open cluster, with its location closely aligned with the cluster member stars; in contrast, J0835$-$4510 and J2337+6151 are positioned at the outer regions of their respective clusters.
Our results for possible associations of pulsars with moving groups are presented in \autoref{append:psr-mg}.

In \autoref{tab:current_psr_oc_pair}, there are four distinct pulsars. Notably, three of these pulsars are associated with supernova remnants (SNRs). 
These include the Vela pulsar J0835$-$4510 \citep{LVM68}, the Crab pulsar J0534+2200 \citep{CCL+69}, and PSR J2337+6151, which is associated with G114.3+0.3 \citep{FRS93, KPH+93}.
Cross matching SNRs with the latest catalog of open clusters might be an interesting topic for further investigations.

\begin{table*}[t]
    \centering
    \setlength{\tabcolsep}{5.3pt}
    \renewcommand{\arraystretch}{1.1}
    \begin{tabular}{ccccccccc}
        \hline\hline
        Name & Prob & $d$ & $\mu_{\alpha} \cos \delta$ & $\mu_{\delta}$ & Age & $P_0$ & $r_{50}$ & $M_{\rm tot}$ \\
         & ($\%$) & (kpc) & ($\operatorname{mas~yr}^{-1}$) & ($\operatorname{mas~yr}^{-1}$) & (yr) & (s) & (deg/pc) & ($\mathrm{M}_{\odot}$) \\
        \midrule 
J2225$+$6535 & -- & $0.83^{+0.17}_{-0.10}$ & $147.2 \pm 0.3$ & $126.5 \pm 0.1$ & $1.12 \times 10^{6}$ & 0.68 & -- & -- \\
HSC~764 & 69.0 (68.7) & $0.865 \pm 0.001$ & $-2.44 \pm 0.01$ & $-4.46 \pm 0.04$ & $1.42 \times 10^{7}$ & -- & 3.05/46.18 & 334.76 \\
OC~0185 & 24.7 (25) & $0.909 \pm 0.001$ & $-1.30 \pm 0.02$ & $-2.49 \pm 0.02$ & $6.44 \times 10^{6}$ & -- & 1.25/19.86 & 593.28 \\
Theia~4069 & 15.6 (13.7) & $1.086 \pm 0.003$ & $2.30 \pm 0.04$ & $-0.20 \pm 0.03$ & $8.21 \times 10^{7}$ & -- & 0.65/12.27 & 244.66 \\
        \midrule 
J0835$-$4510 & \multirow{2}{*}{36.9 (37.1)} & $0.29 \pm 0.02$ & $-49.68 \pm 0.06$ & $29.90 \pm 0.10$ & $1.13 \times 10^{4}$ & 0.09 & -- & -- \\
OC~0450 & & $0.3236 \pm 0.0001$ & $-11.40 \pm 0.02$ & $4.66 \pm 0.02$ & $2.17 \times 10^{7}$ & -- & 3.14/17.78 & 251.32 \\
        \midrule 
J1811$-$2405$^{\dagger}$ & -- & $1.40^{+0.40}_{-0.20}$ & $0.53 \pm 0.06$ & $0 \pm 0$ & $3.06 \times 10^{9}$ & 0.003 & -- & -- \\
Theia~282 & 33.3 (13.1) & $0.784 \pm 0.001$ & $-1.79 \pm 0.02$ & $-1.85 \pm 0.01$ & $9.01 \times 10^{7}$ & -- & 2.21/30.28 & 401.53 \\
NGC~2202 & 19.3 (4.6) & $0.823 \pm 0.002$ & $-0.91 \pm 0.01$ & $-3.77 \pm 0.02$ & $1.15 \times 10^{8}$ & -- & 1.90/27.30 & 228.11 \\
Mamajek~4 & 15.8 (13.3) & $0.4442 \pm 0.0002$ & $3.66 \pm 0.07$ & $-21.67 \pm 0.06$ & $3.12 \times 10^{8}$ & -- & 2.41/18.67 & 702.66 \\
        \midrule 
J0108$+$6608 & \multirow{2}{*}{27.2 (24.4)} & $2.14^{+0.15}_{-0.15}$ & $-32.75 \pm 0.03$ & $35.15 \pm 0.04$ & $1.56 \times 10^{6}$ & 1.28 & -- & -- \\
UBC~1577 & & $2.23 \pm 0.01$ & $-1.34 \pm 0.02$ & $-1.22 \pm 0.01$ & $1.13 \times 10^{8}$ & -- & 0.54/20.89 & 373.29 \\
        \midrule 
J0538$+$2817 & \multirow{2}{*}{26.6 (27.5)} & $1.30^{+0.20}_{-0.20}$ & $-23.57 \pm 0.10$ & $52.87 \pm 0.10$ & $6.18 \times 10^{5}$ & 0.14 & -- & -- \\
CWNU~2136 & & $1.277 \pm 0.006$ & $0.73 \pm 0.02$ & $-5.57 \pm 0.02$ & $3.31 \times 10^{8}$ & -- & 1.99/44.40 & 375.45 \\
        \midrule 
J1757$-$5322$^{\dagger}$ & \multirow{2}{*}{26.2 (18.7)} & $0.80^{+0.25}_{-0.15}$ & $-2.48 \pm 0.08$ & $-10.03 \pm 0.16$ & $5.34 \times 10^{9}$ & 0.01 & -- & -- \\
Mamajek~4 & & $0.4442 \pm 0.0002$ & $3.66 \pm 0.07$ & $-21.67 \pm 0.06$ & $3.12 \times 10^{8}$ & -- & 2.41/18.67 & 702.66 \\
        \midrule 
J0737$-$3039A$^{\dagger}$ & \multirow{2}{*}{23.0 (14.9)} & $1.10^{+0.20}_{-0.10}$ & $-3.8 \pm 0.7$ & $2.1 \pm 0.3$ & $2.04 \times 10^{8}$ & 0.02 & -- & -- \\
OC~0450 & & $0.3236 \pm 0.0001$ & $-11.40 \pm 0.02$ & $4.66 \pm 0.02$ & $2.17 \times 10^{7}$ & -- & 3.14/17.78 & 251.32 \\
        \midrule 
J1456$-$6843 & \multirow{2}{*}{22.4 (38.2)} & $0.43^{+0.06}_{-0.05}$ & $-39.5 \pm 0.4$ & $-12.3 \pm 0.3$ & $4.22 \times 10^{7}$ & 0.26 & -- & -- \\
Theia~181 & & $0.5480 \pm 0.0005$ & $-4.01 \pm 0.02$ & $-5.28 \pm 0.02$ & $1.13 \times 10^{8}$ & -- & 1.07/10.22 & 291.52 \\
        \midrule 
J0147$+$5922 & \multirow{2}{*}{21.3 (29.3)} & $2.02^{+0.46}_{-0.16}$ & $-6.39 \pm 0.09$ & $3.80 \pm 0.08$ & $1.21 \times 10^{7}$ & 0.20 & -- & -- \\
UBC~1577 & & $2.23 \pm 0.01$ & $-1.34 \pm 0.02$ & $-1.22 \pm 0.01$ & $1.13 \times 10^{8}$ & -- & 0.54/20.89 & 373.29 \\
        \midrule 
J1017$-$7156$^{\dagger}$ & \multirow{2}{*}{21.1 (0)} & $0.26^{+0.08}_{-0.08}$ & $-7.410 \pm 0.012$ & $6.871 \pm 0.012$ & $1.67 \times 10^{10}$ & 0.002 & -- & -- \\
HSC~764 & & $0.865 \pm 0.001$ & $-2.44 \pm 0.01$ & $-4.46 \pm 0.04$ & $1.42 \times 10^{7}$ & -- & 3.05/46.18 & 334.76 \\
        \midrule 
J0630$-$2834 & -- & $0.32^{+0.05}_{-0.04}$ & $-46.3 \pm 1.0$ & $21.3 \pm 0.6$ & $2.77 \times 10^{6}$ & 1.24 & -- & -- \\
Collinder~132 & 20.1 (49.7) & $0.637 \pm 0.001$ & $-4.13 \pm 0.02$ & $3.73 \pm 0.02$ & $1.39 \times 10^{7}$ & -- & 0.95/10.61 & 132.50 \\
CWNU~45 & 19.5 (49.6) & $0.6003 \pm 0.0006$ & $-3.14 \pm 0.02$ & $6.28 \pm 0.02$ & $3.31 \times 10^{7}$ & -- & 0.98/10.24 & 266.38 \\
        \midrule 
J0323$+$3944 & \multirow{2}{*}{16.9 (12.1)} & $0.95^{+0.04}_{-0.03}$ & $26.50 \pm 0.05$ & $-30.77 \pm 0.02$ & $7.56 \times 10^{7}$ & 3.03 & -- & -- \\
HSC~764 & & $0.865 \pm 0.001$ & $-2.44 \pm 0.01$ & $-4.46 \pm 0.04$ & $1.42 \times 10^{7}$ & -- & 3.05/46.18 & 334.76 \\
        \midrule 
    \end{tabular}
    \caption{As in \autoref{tab:current_psr_oc_pair}, but for pulsars and potentially associated open clusters in the past. In the association probability column, values outside parentheses are obtained from a uniform pulsar radial velocity distribution in the range of $[-200, 200]\,\mathrm{km\,s^{-1}}$, while those in parentheses assumed Gaussian distribution of $152 \pm 10 \, \mathrm{km \, s^{-1}}$ for normal pulsars or $54 \pm 6 \, \mathrm{km \, s^{-1}}$ for recycled pulsars (names marked with the dagger symbol). Since the two distribution models are very different, it is interesting to note that the results are mostly comparable except the pair of PSR J1017$-$7156 and HSC~764, for which no association can be found for a restrictive Gaussian velocity distribution.}
    \label{tab:past_psr_oc_pair}
\end{table*}

\subsection{Possible Pulsar-Cluster Associations in the Past}
\label{sec:results-past}

By tracing the past trajectories of pulsars and open clusters, we identified 17 possible associations with probability $>15\%$ under the default pulsar radial velocity distribution model (uniform between $-200$ and $200 \, \mathrm{km \, s^{-1}}$), with relevant parameters listed in \autoref{tab:past_psr_oc_pair}.
These 17 associations involve 12 pulsars and 13 clusters, as some pulsars could be associated with multiple clusters, or multiple pulsars could originate from the same cluster.
We note that PSR J0835$-$4510 and OC 0450 are present in both \autoref{tab:current_psr_oc_pair} and \autoref{tab:past_psr_oc_pair}. This is unsurprising because pulsars listed in \autoref{tab:current_psr_oc_pair} are those likely to remain inside the clusters since the start of their lifetimes.
The other three pulsars in \autoref{tab:current_psr_oc_pair} are found to be associated with their clusters in the past with a decent probability of $\sim 10\%$.

In \autoref{tab:past_psr_oc_pair_rv152} of \autoref{append:psr-oc_past}, we present additional 10 candidate pulsar-open cluster associations, i.e., those with probability $>15\%$ under the alternative pulsar radial velocity distribution model.
These 10 associations involve 8 pulsars and 7 clusters that are not present in \autoref{tab:current_psr_oc_pair} or \autoref{tab:past_psr_oc_pair}.
A detailed analysis of all the candidate associations is beyond the scope of this study.
Below we illustrate our calculations with one specific pulsar, J0630$-$2834.
This pulsar is chosen as it is the only one for which the birth place has been investigated in the literature.

\begin{figure*}[!htbp]
    \centering
    \includegraphics[width=0.98\textwidth, trim=0 0 0 0, clip]{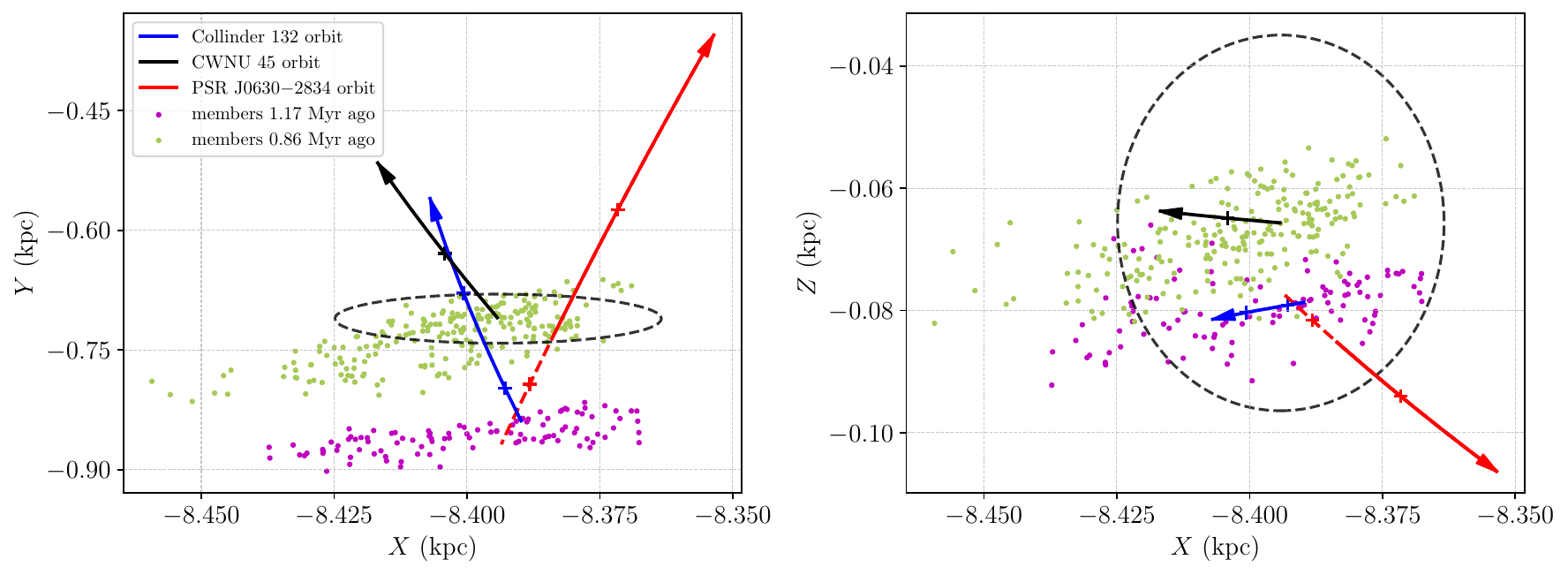}
    \caption{Past trajectories of PSR~J0630$-$2834 and its potential birth open clusters -- Collinder~132 and CWNU~45. The `+' symbol marks a 0.5~Myr interval along the paths.
    Blue (black) lines trace the past trajectories of the clusters, while magenta and green dots indicate the past positions of member stars in Collinder~132 and CWNU~45, respectively. Assuming a radial velocity of $-151~\mathrm{km~s^{-1}}$, the pulsar's trajectory (red line) crossed CWNU~45 about 0.86~Myr ago (solid line) and reached the region of Collinder~132 about 1.17~Myr ago (dashed line). The black dashed circle shows the $3r_{50}$ extent of CWNU~45.
}
    \label{fig:combined_orbits_two_OC}
\end{figure*}

\begin{figure}[htbp]
    \centering
    \includegraphics[width=\linewidth]{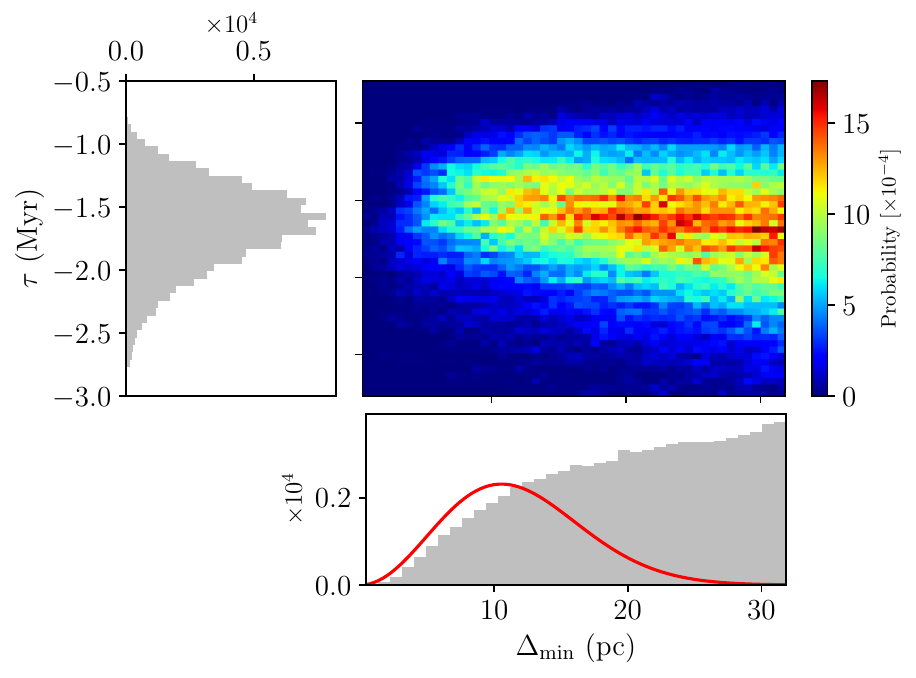}
    \caption{Distributions of minimum separations $\Delta_{\text{min}}$ and corresponding encounter times $\tau$ for PSR J0630$-$2834 and Collinder~132, with an encounter time of $1.6^{+0.4}_{-0.3}$~Myr. 
    The solid red curve is the theoretically expected distribution of $\Delta_{\text{min}}$ (\autoref{eq:2}), while the color bar indicates the probability of each pixel.}
    \label{fig:dist_time_distribution}
\end{figure}

From \autoref{tab:past_psr_oc_pair}, PSR J0630$-$2834 is shown to be associated with two open clusters -- Collinder 132 and CWNU 45 -- with nearly identical probability.
\autoref{fig:combined_orbits_two_OC} illustrates the trajectories of the pulsar and two clusters.
This plot visualizes the two encounter scenarios: $\unit[1.17]{Myr}$ ago the pulsar was inside Collinder 132 and then crossed CWNU 45 later ($\unit[0.86]{Myr}$ ago); in this plot the assumed pulsar radial velocity is $\unit[-151]{km\, s^{-1}}$.
Although we cannot draw a conclusion as to which cluster is more likely to be the pulsar's birth site, we speculate that Collinder 132 is the birth place and CWNU 45 is an accidental crossing along the fly path for two reasons.
First, the separation between the pulsar and cluster center during encounters was smaller for Collinder 132 than CWNU 45; it could be close to zero for the former, whereas the smallest separation is $\approx \unit[10]{pc}$ for the latter.
Second, the pulsar's 3D position was more aligned with member stars of Collinder 132 than those of CWNU 45. \autoref{fig:dist_time_distribution} presents the distributions of minimum separations and corresponding encounter times for PSR J0630$-$2834 and Collinder~132.

In the next section, we present a more in-depth study on the interesting candidates found in \autoref{sec:results-present} and \autoref{sec:results-past}.

\section{Discussions} \label{sec:Discussions}
\subsection{On the origin of PSR J1302$-$6350} \label{sec:disscus_current}

PSR J1302$-$6350 (B1259$–$63), is a gamma-ray pulsar orbiting the massive Be star, LS~2883, with a high eccentricity ($e\approx0.87$) and a long period ($P_{\rm orb}\approx1237\,\mathrm{day}$)  \citep{JML+92, JML+94}. This binary system has been extensively studied in the literature as a rare type of binaries with strong interactions between the pulsar's relativistic wind and the companion star's wind and disk. Several studies proposed that it is likely to be originated in the Cen~OB1 association.
Our analysis suggests that it could be associated with two open clusters, UBC 525 and HSC 2546, with the former being more likely solely based on spatial coincidence calculations.
Below we compare the two scenarios in more detail.

\cite{NRH+11} analyzed the interstellar absorption lines of LS~2883 and derived a distance of $2.3\pm0.4\,\mathrm{kpc}$, which is consistent with the distance to Cen~OB1. 
They argued that PSR J1302$-$6350/LS~2883 may be a member of Cen~OB1. 
Using pulsar timing observations, \cite{SGM14} determined the proper motion of J1302$-$6350 to be ($-6.6\pm1.8, -4.4\pm1.4$)~$\operatorname{mas~yr}^{-1}$. 
They inferred the birth position of the pulsar to be ($l,\,b$) = ($303.9^\circ$, $-0.6^\circ$), which is closer to the center of the Cen~OB1 association than its current location. 
Additionally, \cite{MDS+18} measured the pulsar distance to be $2.6^{+0.4}_{-0.3}\,\mathrm{kpc}$ with VLBI, consistent with the $2.6\pm0.5\,\mathrm{kpc}$ distance of Cen OB1 \citep{CO13}. They suggested that the binary pulsar system was formed within Cen OB1 with a small natal kick, resulting in a displacement of only about $\unit[8]{pc}$ over the past $\sim 3\times10^{5}$~yr; in comparison, the linear size of diameter of Cen OB1 is $\unit[270]{pc}$ \citep{CO13}.

\begin{figure}[htbp]
    \centering
    \includegraphics[width=\linewidth]{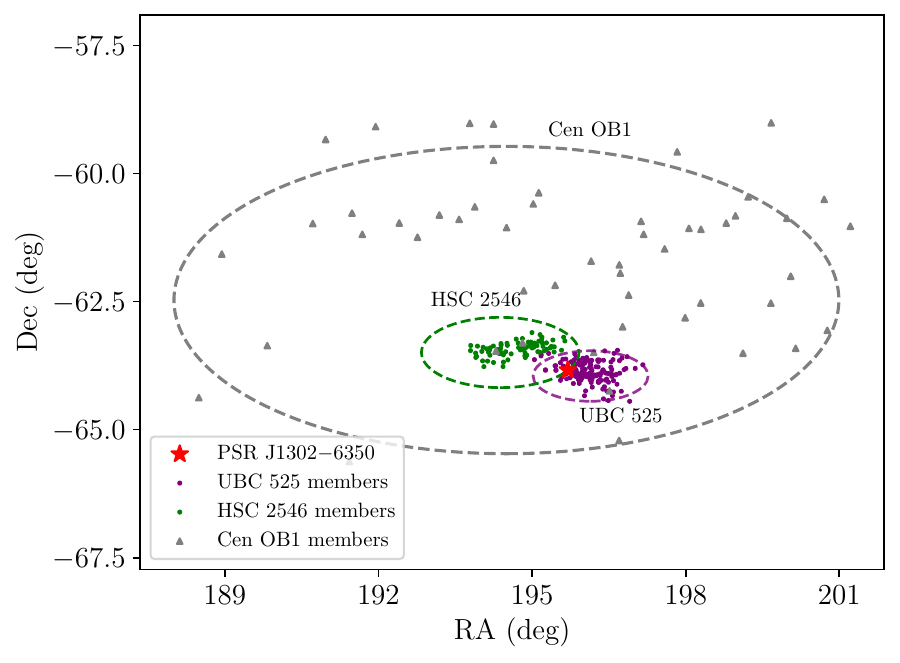}
    \caption{The sky location of PSR J1302$-$6350, two open clusters (UBC~525 and HSC~2546), the stellar association Cen OB1 and their member stars. The magenta and green dashed ellipses mark the $3r_{50}$ regions of UBC 525 and HSC 2546, respectively, with their member stars shown as dots. The gray dashed ellipse indicates the region of Cen OB1 with a radius of $3^\circ$, and the gray triangles denote its member stars \citep{CO13}.}
    \label{fig:Cen_OB1_coord}
\end{figure}

 
In \autoref{fig:Cen_OB1_coord}, we show the positions of the open clusters UBC~525 and HSC~2546, and Cen OB1.
First, we note that both UBC~525 and HSC~2546 lie within the boundaries of Cen OB1, although the actual distances are different.
Adopting the Gaia DR2 distance to the pulsar \citep{JKC+18}, we derive a present-day spatial association probability of 10\% with UBC~525, slightly higher than that with Cen OB1 (9\%), while the probability for HSC 2546 is the lowest, at only 1\%.

We disregard the association probability with HSC~2546 for the following two reasons.
First, as can be seen in \autoref{fig:Cen_OB1_coord}, the binary pulsar is separated from member stars of HSC 2546, whereas it is located near the center of UBC 525.
Second, the mass of the companion LS~2883 is estimated to be around $15-\unit[31]{{M}_\odot}$ \citep{MDS+18}; this is consistent with the high-mass end of member stars in UBC 525, which has a total mass of $\approx \unit[1457]{{M}_{\odot}}$ and comprises 173 member stars, including 12 stars with masses exceeding $\unit[8]{{M}_\odot}$ and the most massive star being $\approx \unit[18]{{M}_\odot}$ \citep{HR24}.
On the other hand, HSC~2546 hosts only two massive stars, with masses of $\sim\unit[8]{{M}_\odot}$ and $\sim \unit[9]{{M}_\odot}$, respectively.

\begin{figure*}[htbp]
    \centering
    \includegraphics[width= 0.8\linewidth]{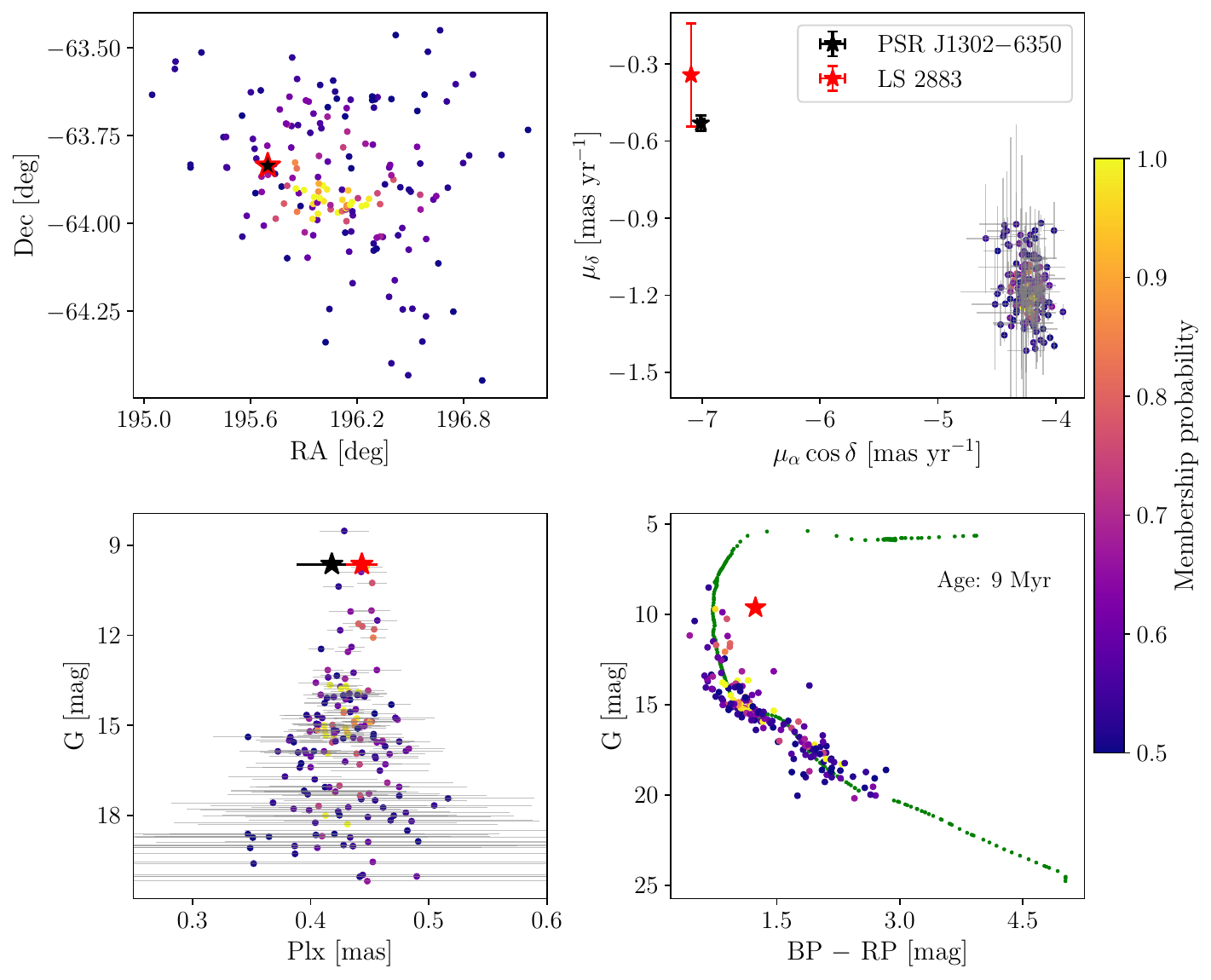}
    \caption{Properties of the member stars of UBC~525 and the binary system PSR J1302$-$6350/LS 2883. In all panels, the color of the dot points indicates the probability of cluster membership (as defined in the colorbar); the black (red) star marks PSR J1302$-$6350 (LS~2883). In the upper-right panel, the pulsar proper motion was measured in \citet{MDS+18}. The lower-left panel shows parallax versus G magnitude, with the black and red stars indicating the parallax ranges of LS~2883 using Gaia DR2 \citep{JKC+18} and DR3, respectively. The lower-right panel shows the color-magnitude diagram using Gaia DR3's photometric bands. All data are taken from Gaia DR3 unless otherwise specified.}
    \label{fig:J1302_UBC525_Plx}
\end{figure*}

In \autoref{fig:J1302_UBC525_Plx}, we plot the sky location, proper motion and parallax for the member stars of UBC 525, as well as PSR J1302$-$6350 and its companion star LS~2883.
Consistency is observed for all parameters except the proper motion, which can be attributed to the natal kick imparted to the pulsar during supernova explosion.
The ages of PSR J1302$-$6350 ($\sim 0.3\,\mathrm{Myr}$) and UBC~525 ($\sim9\,\mathrm{Myr}$) are also consistent, with the difference being the lifetime of the pulsar progenitor star.

In the lower right panel of \autoref{fig:J1302_UBC525_Plx}, we present a color-magnitude diagram of UBC 525 using photometric data from Gaia DR3. The corresponding isochrone\footnote{The isochrone is queried using a script written by Zhaozhou Li (as part of \citealt{Li2020a}), \url{https://github.com/syrte/query_isochrone}.} is based on parameters derived by \citet{Cavallo2024}: $\log {\rm Age} = 6.97$, distance modulus = 11.59, extinction $A_V=2.38$, metallicity [Fe/H]=$-0.03$ \citep{Li2025}. Specifically, we adopt PARSEC isochrones \citep{Nguyen2022} with the Gaia EDR3 photometric system \citep{Riello2021} and the bolometric correction model developed by \citet{Chen2019c}.
By fitting the G-band magnitude to the corresponding isochrone, we derive a mass estimate of $\unit[16.78] {M_\odot}$ for LS 2883.

In summary, although we can not rule out the possibility of the binary pulsar J1302$-$6350 originated and still reside in the Cen OB1 association, we find compelling evidence for it being part of the open cluster UBC 525.
In this sense, UBC~525 is an interesting target for pulsar search with sensitive radio telescopes, and new pulsar discoveries will confirm its association with PSR~J1302$-$6350.

\subsection{On the origin of PSR J0630$-$2834} \label{sec:disscus_past}

\begin{table}[t]
\centering
\caption{Association probability between PSR J0630$-$2834 and the open cluster Collinder 132 or the Antlia SNR for different radial velocity distributions. For Gaussian distributions, we assume that there is equal likelihood for the radial velocity being positive or negative.}
\label{tab:0630_prob}
\begin{tabular}{cccc}
\toprule
\multirow{2}{*}{Distribution} & \multirow{2}{*}{RV ($\mathrm{km\,s^{-1}}$)} & \multicolumn{2}{c}{Association Probability (\%)} \\
\cmidrule(lr){3-4}
 & & Collinder 132 & Antlia SNR \\
\midrule
\multirow{6}{*}{Gaussian} & 152$\pm$10 & 50 & 0 \\
                       & 313$\pm$28 & 1 & 0.01 \\
                       & 246$\pm$22 & 29 & 0 \\
                       & 133$\pm$8 & 44 & 0 \\
                       & 200$\pm$100 & 0.05 & 0.1 \\
                       & 50$\pm$10 & 0 & 0 \\
\midrule
\multirow{4}{*}{Uniform} & [$-$200, 200] & 32 & 0 \\
                         & [0, 200] & 0 & 0 \\
                         & [0, 500] & 0 & 1 \\
                         & [300, 500] & 0 & 3 \\
\bottomrule
\end{tabular}
\end{table}

\begin{figure}[htbp]
    \centering
    \includegraphics[width=\linewidth]
    {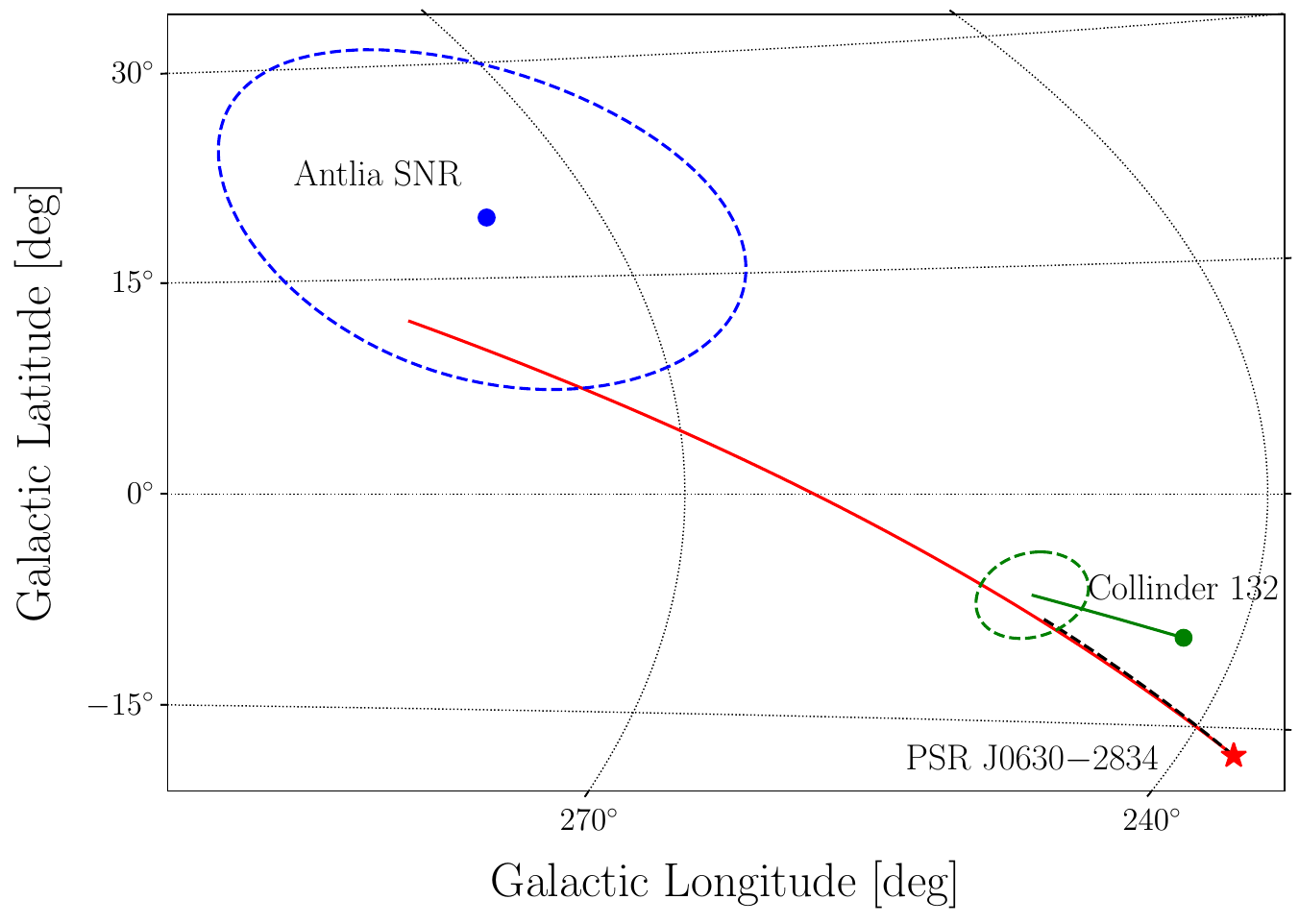}
    \caption{The past trajectory of PSR J0630$-$2834 (with its current position marked by a red star) relative to the Antlia SNR and the open cluster Collinder 132. The blue dot indicates the position of Antlia SNR, with its extension marked by a blue dashed circle. The red solid line shows the pulsar trajectory up to 0.72 Myr ago for a radial velocity of $454~\mathrm{km\,s^{-1}}$. The black dashed line assumes a pulsar radial velocity of $-230~\mathrm{km\,s^{-1}}$ and shows the pulsar trajectory up to 1.06 Myr ago which coincide with Collinder 132, whose extent is shown by a green dashed circle and current position by a green dot.}
    \label{fig:Antlia_Col132_J0630_GlGb}
\end{figure}

Previous studies have proposed an association between PSR J0630$-$2834 (B0628$-$28) and the Antlia SNR.
\cite{TTN+13} suggested that they were born from the same supernova event and that the runaway star HIP 47155 was a former companion to PSR~J0630$-$2834.
Under this hypothesis, they estimated an SNR age of approximately 1.2 Myr.
Since \cite{TTN+13} did not specify Antlia's 3D velocity used in their analysis, and such data are unavailable in the literature, we were unable to reproduce their results.
Nevertheless, we can compare the SNR-association hypothesis with an open cluster origin quantitatively.
The Antlia SNR is located at Galactic coordinates ($l,\,b$) = ($276.5^\circ$, $19^\circ$) with a large projected diameter of $12^\circ$, and the distance is likely to be $\lesssim 500$~pc.
The predicted supernova position 1.2 Myr ago found in \cite{TTN+13} is ($l,\,b$) = ($270^\circ$, $19^\circ$), which is close to the current SNR position given the large angular size of the SNR.
We can compute the probability of PSR J0630$-$2834 crossing the SNR within the past $\sim$ Myr for different radial velocity distributions.

In \autoref{tab:0630_prob}, we list the probabilities of a close encounter between PSR J0630$-$2834 and the open cluster Collinder 132 or the Antlia SNR.
For a pulsar radial velocity of $200\pm 100~\mathrm{km\,s^{-1}}$, it is indeed possible that PSR J0630$-$2834 was born in the Antlia SNR, although with a very low probability at $0.1\%$.
In contrast, the probability for PSR J0630$-$2834 originating from Collinder 132 is significantly higher; using a Gaussian velocity distribution of $150\pm 10~\mathrm{km\,s^{-1}}$ gives $50\%$ chance.
Since we assume there is equal likelihood for the pulsar radial velocity being positive or negative, it is interesting to note that the half chance for PSR J0630$-$2834 associated with Collinder 132 is entirely due to negative radial velocity (moving toward the Earth).
On the other hand, for the pulsar to be born in the Antlia SNR, the pulsar radial velocity has to be positive.
\autoref{fig:Antlia_Col132_J0630_GlGb} illustrates the pulsar trajectory and sky location of the Antlia SNR as well as Collinder 132 in the Galactic coordinate system.
It is possible to trace the pulsar to the SNR or to the open cluster, depending on the direction of pulsar's radial velocity.

While our probabilistic trajectory calculations indicate that PSR J0630$-$2834 is more likely associated with Collinder 132, new SNR age estimate rules out the pulsar association with Antlia.
\cite{Fesen21AntliaSNR} inferred the age of Antlia SNR to be $\lesssim 10^{5}$~yr based on shock velocity measurements derived from optical spectra; the age is an order of magnitude smaller than the $\sim 1$ Myr required fly time.

\section{Conclusions} \label{sec:Conclusions}

Stars are predominately born in clusters from giant molecular clouds.
The massive stellar progenitors of neutron stars are expected to be part of star clusters in the early stage.
Most clusters may dissolve well before their high-mass member stars explode and form neutron stars. In case neutron stars are formed in the cluster, they may also be kicked out of the cluster during supernova explosion.
Neutron stars in open clusters are therefore a rare specie, and if they do exist, they would provide unique insights into the formation of neutron stars and stellar evolution.

In this work, we performed the first census of pulsars that could be associated with open clusters by cross matching the known pulsar population with thousands of open clusters identified by the Gaia mission.
We obtained a list of pulsar candidates, namely, 4 pulsars that could currently be part of an open cluster, and another 19 pulsars that could be born from an open cluster.
Taken at face values, the probability for pulsars to be born with open clusters and remain inside the cluster is indeed very low; $\lesssim 4$ out of 164 from our census belong to this category. In contrast, 19 out of 145 pulsars are likely to be born in open clusters; this fraction is higher than that of globular-cluster pulsars\footnote{Since only preliminary statements can be made regarding the pulsar population originating from Galactic open clusters, we have ignored any selection effects present in the adopted catalogs of pulsars and open clusters.}.
Making a definitive association of pulsars with open clusters is very challenging because of the large pulsar distance uncertainties and the unconstrained radial velocity distribution.
Therefore our candidates are presented as interesting targets for future studies.

The first highlight from this work is that the identification of Collinder 132 as the possible birth place of PSR~J0630$-$2834.
Our calculations of the past trajectory of PSR J0630$-$2834 indicate that it is more likely to be born in Collinder 132 ($\unit[1.6]{Myr}$ ago) instead of the Antlia SNR as suggested in previous studies; incorporating the new SNR age estimate rules out the SNR association completely.

We find compelling evidence for the association of the gamma-ray binary pulsar J1302$-$6350 with the open cluster UBC 525.
This is made possible by Gaia measurements of its companion stars LS 2883; as a side output from this work, we update the pulsar distance to be $2.26\pm 0.07$~kpc and measure the companion mass to be $\unit[16.8]{ M_\odot}$.
Our study underscores the importance of searching for radio pulsars in open clusters using the most sensitive radio telescopes such as FAST, MeerKAT and the Square Kilometer Array.

\begin{acknowledgments}
This work is supported by the National Natural Science Foundation of China (Grant No.~12203004), the National Key Research and Development Program of China (No. 2023YFC2206704), the Fundamental Research Funds for the Central Universities, and the Supplemental Funds for Major Scientific Research Projects of Beijing Normal University (Zhuhai) under Project ZHPT2025001. We thank the anonymous referee for very useful comments. Li, L. thanks the support of NSFC No. 12303026 and the Young Data Scientist Project of the National Astronomical Data Center. You, Z.-Q. is supported by the National Natural Science Foundation of China (Grant No.~12305059); The Startup Research Fund of Henan Academy of Sciences (No.~241841224); The Scientific and Technological Research Project of Henan Academy of Science (No.~20252345003); Joint Fund of Henan Province Science and Technology R\&D Program (No.~235200810111). 
Part of the analyses were performed on the OzSTAR national facility at Swinburne University of Technology. The OzSTAR program receives funding in part from the Astronomy National Collaborative Research Infrastructure Strategy (NCRIS) allocation provided by the Australian Government, and from the Victorian Higher Education State Investment Fund (VHESIF) provided by the Victorian Government. 

\end{acknowledgments}

\software{astropy \citep{Astropy13, Astropy18, Astropy22}, Gala \citep{P17}, Matplotlib \citep{Hunter07}, Numpy \citep{HMv+20}.}

\newpage
\appendix

\section{Independent distance estimates of pulsars}
\label{appendix_distance}

In \autoref{tab:pulsar_independent_dist}, we list 164 pulsar distance estimates that are independent from the Galactic electron density models.

\startlongtable
\begin{deluxetable*}{ccccccccc}
\centering
\tablecaption{%
Pulsar distance measurements as categorized by measurement methods. References in the table are: (1) \protect\cite{YMW17}, (2) \protect\cite{VWC+12}, 
(3) \protect\cite{SBF+24}, (4) \protect\cite{DDS+23}, (5) \protect\cite{DGB+19}, (6) \protect\cite{BTG+03}, (7) \protect\cite{BBG+02}, (8) \protect\cite{DLR+03}, (9) \protect\cite{LPM+16}, (10) \protect\cite{NBB+14}, (11) \protect\cite{JKC+18}, (12) \protect\cite{GSL+16}, (13) \protect\cite{VK07}, (14) \protect\cite{VBC+09}, (15) \protect\cite{RBS+24}, (16) \protect\cite{TPC+11}.\label{tab:pulsar_independent_dist} }
\tablehead{
\colhead{Pulsar} & \colhead{$D$ (kpc)} & \colhead{Ref.} & 
\colhead{Pulsar} & \colhead{$D$ (kpc)} & \colhead{Ref.} & 
\colhead{Pulsar} & \colhead{$D$ (kpc)} & \colhead{Ref.}
}
\startdata
        &            &                         &     & Parallax   &    & \\
        \hline
        J0030+0451   &$0.329^{+0.006}_{-0.005}$& (4) & J1022+1001   &$0.72^{+0.01}_{-0.02}$    & (5) & J1840+5640   &$1.52^{+0.02}_{-0.14}$    & (5) \\
        J0034$-$0721 &$1.03\pm0.08$            & (2) & J1023+0038   &$1.367^{+0.043}_{-0.039}$ & (1) & J1853+1303   &$2.0\pm0.3$               & (4) \\
        J0040+5716   &$9.77^{+3.13}_{-3.23}$   & (5) & J1024$-$0719 &$1.08\pm0.06$             & (4) & J1856$-$3754 &$0.161^{+0.018}_{-0.014}$ & (13) \\
        J0055+5117   &$2.87^{+0.54}_{-0.39}$   & (5) & J1125$-$6014 &$0.86^{+0.22}_{-0.16}$    & (3) & J1857+0943   &$0.91\pm0.17$             & (14) \\
        J0102+6537   &$2.51^{+0.32}_{-0.25}$   & (5) & J1136+1551   &$0.35\pm0.02$             & (2) & J1901$-$0906 &$1.96^{+0.17}_{-0.23}$    & (5) \\
        J0108+6608   &$2.14\pm0.15$            & (5) & J1239+2453   &$0.84^{+0.06}_{-0.05}$    & (2) & J1909$-$3744 &$1.26\pm0.03$             & (2) \\
        J0125$-$2327 &$1.19^{+0.17}_{-0.13}$   & (3) & J1321+8323   &$1.03^{+0.17}_{-0.04}$    & (5) & J1910+1256   &$3.4^{+1.0}_{-0.6}$       & (4) \\
        J0139+5814	 &$2.6^{+0.3}_{-0.2}$      & (2) & J1455$-$3330 &$1.01\pm0.22$             & (12)& J1913+1400   &$5.42^{+0.75}_{-0.70}$    & (5) \\
        J0147+5922   &$2.02^{+0.46}_{-0.16}$   & (5) & J1456$-$6843 &$0.43^{+0.06}_{-0.05}$    & (2) & J1918$-$0642 &$0.909^{+0.202}_{-0.140}$ & (1) \\
        J0157+6212   &$1.80^{+0.08}_{-0.12}$   & (5) & J1509+5531   &$2.1\pm0.1$               & (2) & J1932+1059   &$0.31^{+0.09}_{-0.06}$    & (2) \\
        J0218+4232   &$3.15^{+0.85}_{-0.60}$   & (1) & J1518+4904   &$0.81\pm0.02$             & (4) & J1937+2544   &$3.15^{+0.32}_{-0.28}$    & (5) \\
        J0323+3944   &$0.80^{+0.04}_{-0.03}$   & (5) & J1532+2745   &$1.60^{+0.29}_{-0.07}$    & (5) & J1939+2134   &$2.9^{+0.3}_{-0.2}$       & (4) \\
        J0332+5434   &$1.68^{+0.07}_{-0.06}$   & (5) & J1537+1155   &$0.94^{+0.07}_{-0.06}$    & (4) & J1946$-$5403 &$1.23^{+0.18}_{-0.14}$    & (3) \\
        J0335+4555   &$2.44^{+0.18}_{-0.12}$   & (5) & J1543+0929   &$5.9^{+0.6}_{-0.5}$       & (2) & J2006$-$0807 &$2.36^{+0.73}_{-0.06}$    & (5) \\
        J0358+5413   &$1.0^{+0.2}_{-0.1}$      & (2) & J1543$-$0620 &$3.11^{+0.51}_{-0.25}$    & (5) & J2010$-$1323 &$2.07^{+0.68}_{-0.53}$    & (5) \\
        J0437$-$4715 &$0.15696\pm0.00011$	   & (15)& J1559$-$4438 &$2.3^{+0.5}_{-0.3}$       & (2) & J2017+0603   &$1.56\pm0.31$             & (1) \\
        J0454+5543   &$1.18^{+0.07}_{-0.05}$   & (2) & J1600$-$3053 &$2.0^{+0.5}_{-0.3}$       & (3) & J2018+2839   &$0.95\pm0.09$             & (7) \\
        J0538+2817   &$1.3\pm0.2$              & (2) & J1603$-$7202 &$0.53^{+0.04}_{-0.16}$    & (1) & J2022+5154   &$1.8^{+0.3}_{-0.2}$       & (2) \\
        J0601$-$0527 &$2.09^{+0.22}_{-0.16}$   & (5) & J1607$-$0032 &$1.07^{+0.06}_{-0.03}$    & (5) & J2046+1540   &$3.22^{+1.04}_{-0.68}$    & (5) \\
        J0610$-$2100 &$1.5^{+0.3}_{-0.2}$      & (4) & J1614$-$2230 &$0.77\pm0.05$             & (12)& J2048$-$1616 &$0.95^{+0.02}_{-0.03}$    & (2) \\
        J0614+2229   &$3.55^{+0.44}_{-0.26}$   & (5) & J1623$-$0908 &$1.71^{+0.34}_{-0.25}$    & (5) & J2055+3630   &$5.0^{+0.8}_{-0.6}$       & (2) \\
        J0621+1002   &$1.6^{+0.5}_{-0.3}$      & (4) & J1629$-$6902 &$1.1^{+0.3}_{-0.2}$       & (3) & J2113+2754   &$1.42\pm0.04$             & (5) \\
        J0629+2415   &$3.00^{+0.57}_{-0.29}$   & (5) & J1640+2224   &$1.39^{+0.13}_{-0.11}$    & (4) & J2113+4644   &$2.20^{+0.36}_{-0.32}$    & (5) \\
        J0630$-$2834 & $0.32^{+0.05}_{-0.04}$  & (2) & J1643$-$1224 &$0.39^{+0.11}_{-0.07}$    & (3) & J2124$-$3358 &$0.48\pm0.04$	           & (3) \\
        J0636+5129   &$0.204^{+0.029}_{-0.022}$& (1) & J1645$-$0317 &$3.97^{+0.33}_{-0.39}$    & (5) & J2129$-$0429 &$2.06^{+0.67}_{-0.21}$    & (11) \\
        J0636$-$3044 &$0.23^{+0.04}_{-0.03}$   & (3) & J1658$-$5324 &$0.75^{+0.17}_{-0.12}$    & (3) & J2144$-$3933 &$0.16^{+0.02}_{-0.01}$    & (2) \\
        J0645+5158   &$0.769^{+0.231}_{-0.144}$& (1) & J1703$-$1846 &$2.88^{+0.45}_{-0.36}$    & (5) & J2145$-$0750 &$0.62^{+0.00}_{-0.02}$    & (5) \\
        J0659+1414   &$0.288^{+0.033}_{-0.027}$& (6) & J1713+0747   &$1.05^{+0.06}_{-0.05}$    & (2) & J2149+6329   &$2.81^{+0.58}_{-0.47}$    & (5) \\
        J0729$-$1836 &$2.04^{+0.39}_{-0.34}$   & (5) & J1723$-$2837 &$0.90^{+0.05}_{-0.04}$    &(11) & J2157+4017   &$2.9^{+0.5}_{-0.4}$       & (2) \\
        J0737$-$3039A&$1.1^{+0.2}_{-0.1}$      & (2) & J1730$-$2304 &$0.44^{+0.05}_{-0.04}$    & (3) & J2214+3000   &$1.000^{+0.111}_{-0.091}$ & (1) \\
        J0814+7429   &$0.433\pm0.008$          & (7) & J1735$-$0724 &$6.68^{+2.03}_{-1.43}$    & (5) & J2222$-$0137 &$0.267\pm0.001$           & (1) \\
        J0820$-$1350 &$1.9\pm0.1$              & (2) & J1738+0333   &$1.47^{+0.12}_{-0.10}$    & (1) & J2225+6535   &$0.83^{+0.17}_{-0.10}$    & (5) \\
        J0823+0159   &$2.66^{+0.60}_{-0.68}$   & (5) & J1741+1351   &$1.075^{+0.061}_{-0.055}$ & (1) & J2241$-$5236 &$1.10^{+0.07}_{-0.06}$    & (3) \\
        J0826+2637    &$0.50\pm0.00$           & (5) & J1741$-$0840 &$3.58^{+0.94}_{-0.55}$    & (5) & J2305+3100   &$4.47^{+0.65}_{-0.58}$    & (5) \\
        J0835$-$4510 &$0.287^{+0.019}_{-0.017}$& (8)& J1744$-$1134 &$0.42\pm0.02$             & (2) & J2313+4253   &$1.06^{+0.08}_{-0.06}$    & (2) \\
        J0837+0610   &$0.62\pm0.06$            & (9)& J1754+5201   &$6.27^{+1.03}_{-0.98}$    & (5) & J2317+2149   &$1.96^{+0.21}_{-0.20}$    & (5) \\
        J0922+0638   &$1.1^{+0.2}_{-0.1}$      & (2) & J1757$-$5322 &$0.80^{+0.25}_{-0.15}$    & (3) & J2322+2057   &$0.80^{+0.21}_{-0.15}$    & (3) \\
        J0953+0755   &$0.261\pm0.005$          & (2) & J1811$-$2405 &$1.4^{+0.4}_{-0.2}$       & (3) & J2322$-$2650 &$0.80^{+0.18}_{-0.12}$    & (3) \\
        J1012+5307    &$0.877\pm0.035$         & (4) & J1820$-$0427 &$2.85^{+0.52}_{-0.35}$    & (5) & J2346$-$0609 &$3.64^{+0.55}_{-0.26}$    & (5) \\
        J1017$-$7156 &$0.256\pm0.079$          & (10)& J1833$-$0338 &$2.45^{+0.48}_{-0.27}$    & (5) & J2354+6155   &$2.42^{+0.28}_{-0.17}$    & (5) \\
         &     \        &          \           & \ & HI absorption & & \\
        \hline
        J0141+6009   &$2.3\pm0.7$         & (2) & J1832$-$0827 &$5.2^{+0.5}_{-0.4}$  & (2) & J1906+0641   &$7\pm2$             & (2) \\
        J1056$-$6258 &$2.4\pm0.5$         & (2) & J1833$-$0827 &$4.5\pm0.5$	         & (2) & J1909+1102   &$4.8^{+1.1}_{-0.8}$ & (2) \\
        J1600$-$5044 &$6.9^{+1.9}_{-0.9}$ & (2) & J1848$-$0123 &$4.4\pm0.4$	         & (2) & J1915+1009   &$7\pm2$             & (2) \\
        J1602$-$5100 &$8.0^{+0.9}_{-0.7}$ & (2) & J1852+0031   &$8\pm2$              & (2) & J1916+1312   &$4.5^{+1.2}_{-0.9}$ & (2) \\
        J1644$-$4559 &$4.5\pm0.4$         & (2) & J1857+0212   &$8\pm2$	             & (2) & J1917+1353   &$5\pm1$	           & (2) \\
        J1709$-$4429 &$2.6^{+0.5}_{-0.6}$ & (2) & J1901+0331   &$7\pm2$              & (2) & J1932+2220   &$10.9^{+1.3}_{-0.8}$& (2) \\
        J1721$-$3532 &$4.6\pm0.6$         & (2) & J1901+0716   &$3.4^{+0.9}_{-0.7}$  & (2) & J2004+3137   &$8^{+2}_{-1}$       & (2) \\
        J1801$-$2304 &$4\pm1$             & (2) & J1902+0556   &$3.6^{+0.6}_{-0.5}$  & (2) & J2257+5909   &$4.1\pm0.7$         & (2) \\
        J1809$-$1943 &$3.6\pm0.5$	      & (1) & J1903+0135   &$3.3^{+0.6}_{-0.5}$  & (2)  \\
         &     \        &          \               & \ & Nebular associations & & \\
        \hline
        J0205+6449   &$3.2\pm0.6$	     & (1) & J1513$-$5908 &$4.4^{+1.3}_{-0.8}$ & (1) & J1856+0113   &$3.3\pm0.6$ & (1) \\
        J0248+6021   &$2.0\pm0.2$        & (16)& J1550$-$5418 &$4.0\pm0.5$	       & (1) & J2229+6114   &$3.0\pm0.6$ & (1) \\
        J0534+2200   &$2.0\pm0.5$        & (1) & J1745$-$2900 &$8.3\pm0.3$	       & (1) & J2337+6151   &$0.7\pm0.1$ & (1) \\
        J1119$-$6127 &$8.4\pm0.4$	     & (1) & J1803$-$2137 &$4.4^{+0.5}_{-0.6}$ & (1) \\
        J1400$-$6325 &$7\pm2$	         & (1) & J1833$-$1034 &$4.1\pm0.3$	       & (1) \\
         &     \        &          \               & \ & Stellar companions & & \\
        \hline
        J0348+0432   &$2.1\pm0.2$	    & (1) & J1302$-$6350 &$2.21^{+0.19}_{-0.12}$ &(11) & J2032+4127   &$1.38^{+0.07}_{-0.06}$  &(11) \\
        J1231$-$1411 &$0.43\pm0.08$     & (1) & J1903+0327   &$7.0\pm1.0$            & (1) \\
\enddata
\end{deluxetable*}

\section{Visualization of the spatial coincidence between pulsars and open clusters}
\label{append:psr-oc}

\begin{figure*}[htbp]
    \centering
    \begin{subfigure}{0.45\textwidth}
        \includegraphics[width=\linewidth]{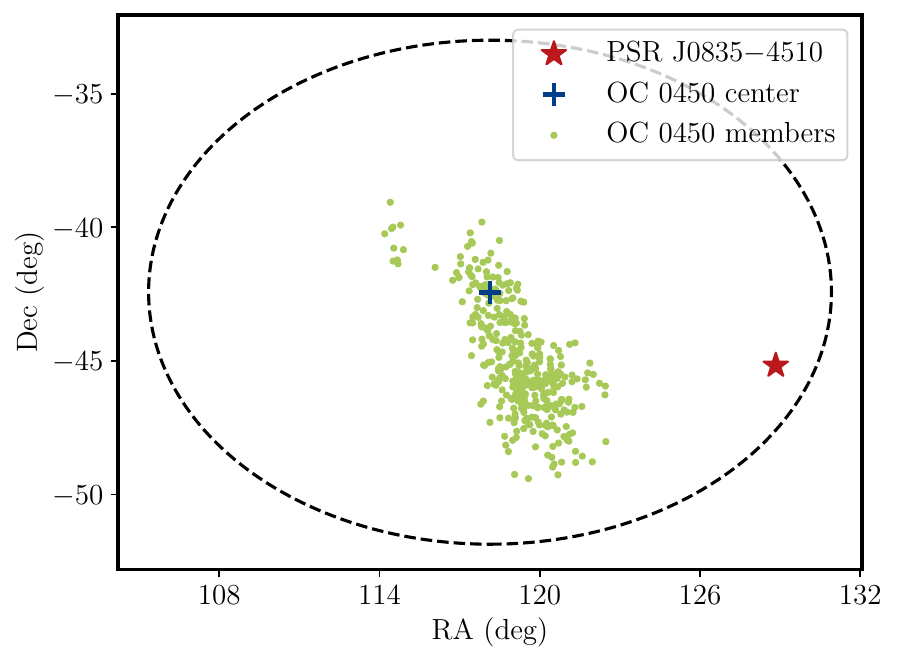}
    \end{subfigure}
    \begin{subfigure}{0.45\textwidth}
        \includegraphics[width=\linewidth, trim=50 162 0 150, clip]{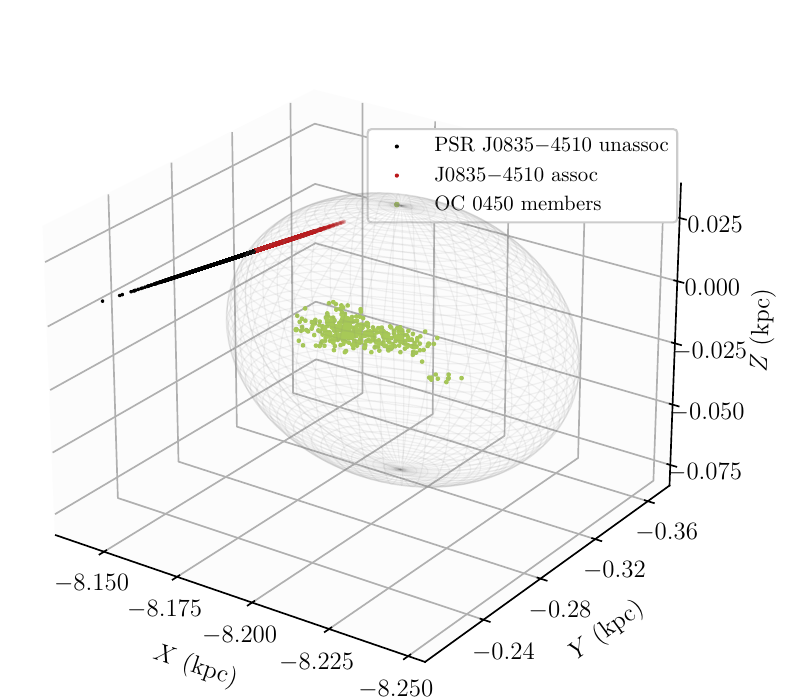}
    \end{subfigure}
    \begin{subfigure}{0.45\textwidth}
        \includegraphics[width=\linewidth]{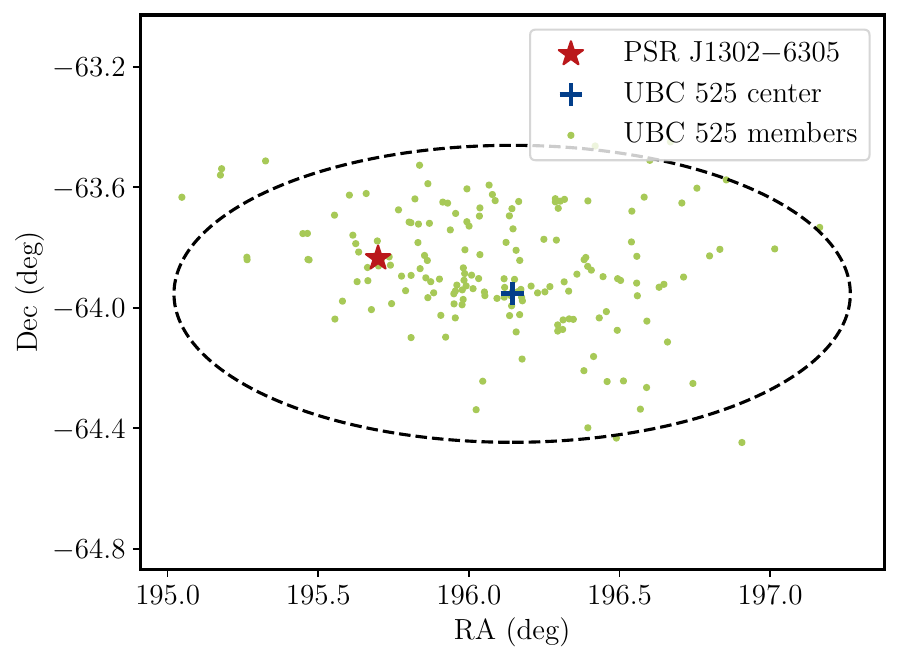}
    \end{subfigure}
    \begin{subfigure}{0.45\textwidth}
        \includegraphics[width=\linewidth, trim=50 162 0 150, clip]{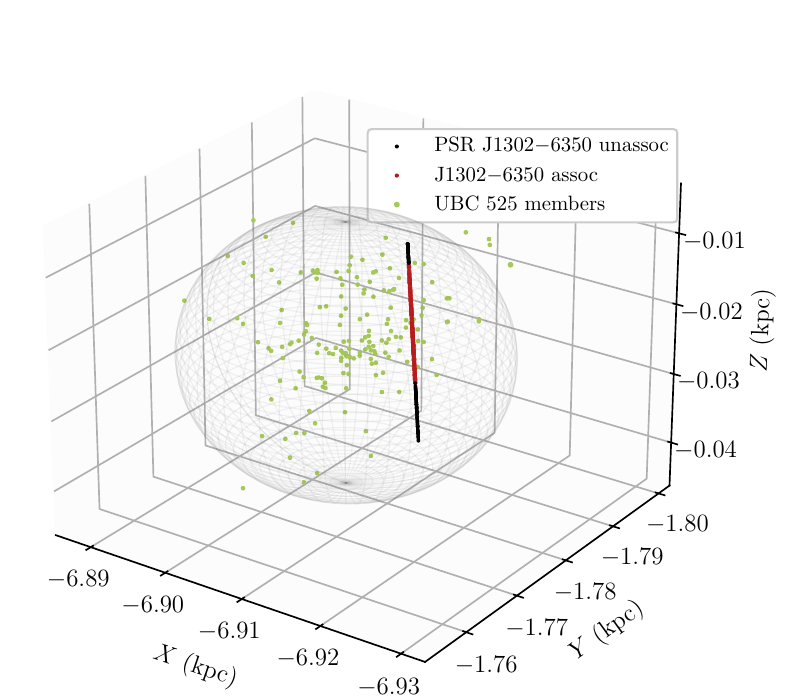}
    \end{subfigure}
    \begin{subfigure}{0.45\textwidth}
        \includegraphics[width=\linewidth]{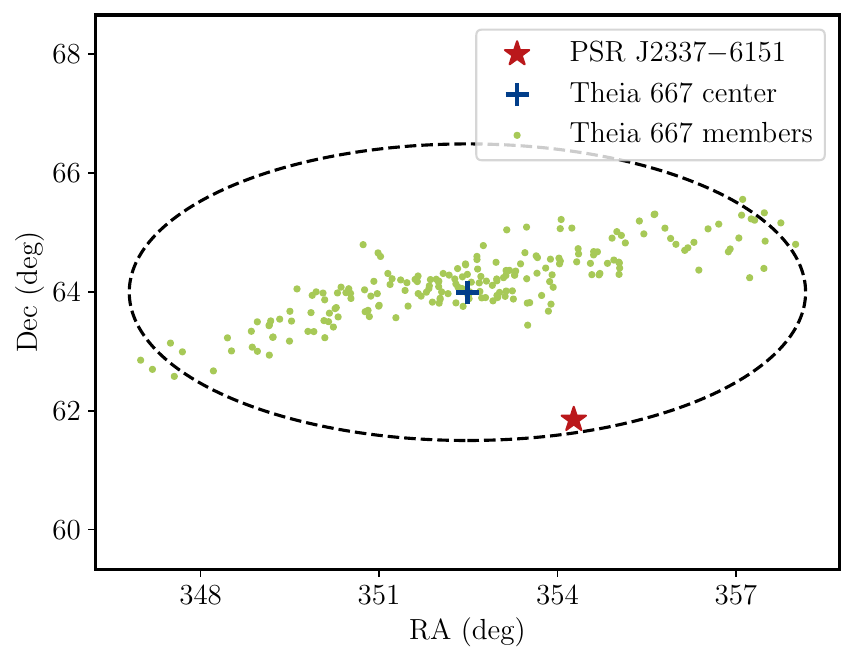}
    \end{subfigure}
    \begin{subfigure}{0.45\textwidth}
        \includegraphics[width=\linewidth, trim=50 162 0 150, clip]{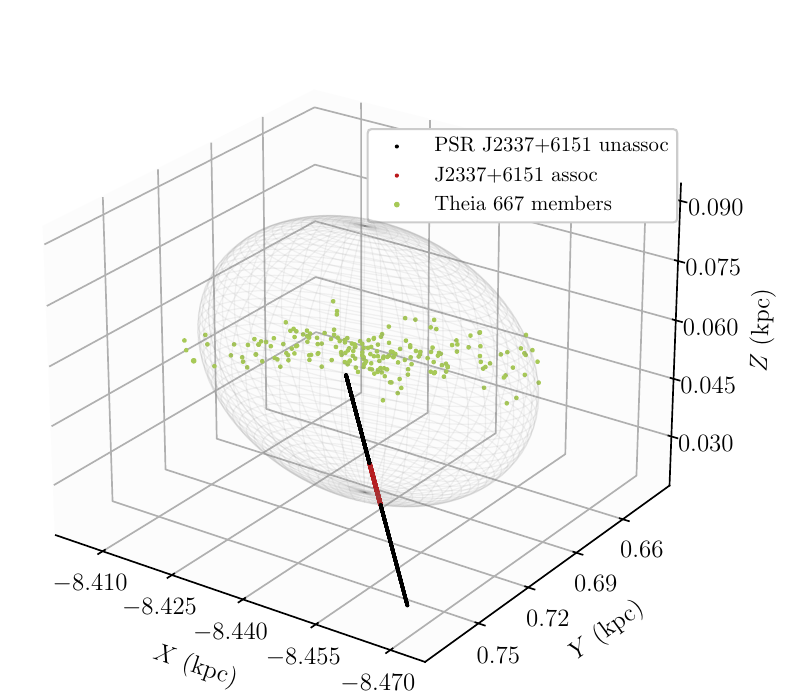}
    \end{subfigure}
    \caption{
    The spatial distribution of three possible pulsar-open cluster associations. 
    The left panels display the sky location for pulsars (red stars), the center of open clusters (blue crosses), and cluster member stars (green points). 
    Each open cluster is shown with a radius of $3r_{50}$ (the black dashed line). The right panels present three-dimensional Cartesian coordinates of pulsars (black dots). The red dots represent those within $3r_{50}$ from the center of the open cluster, while the green points show the cluster members assuming the same distance as cluster center (but different RA and Dec).}
     \label{fig:current_psr_oc_coord}
\end{figure*}

In \autoref{fig:current_psr_oc_coord}, we show the 2D and 3D spatial overlap of the top three pulsar-cluster pairs listed in \autoref{tab:current_psr_oc_pair}.
Note that in the 3D plot, the distances of cluster member stars are assumed to be the same as the cluster center; if their distances uncertainties are taken into account, the spread would be more significant.

\begin{figure*}[htp]
    \centering
    \includegraphics[width=0.46 \textwidth, trim=0 0 0 0, clip]{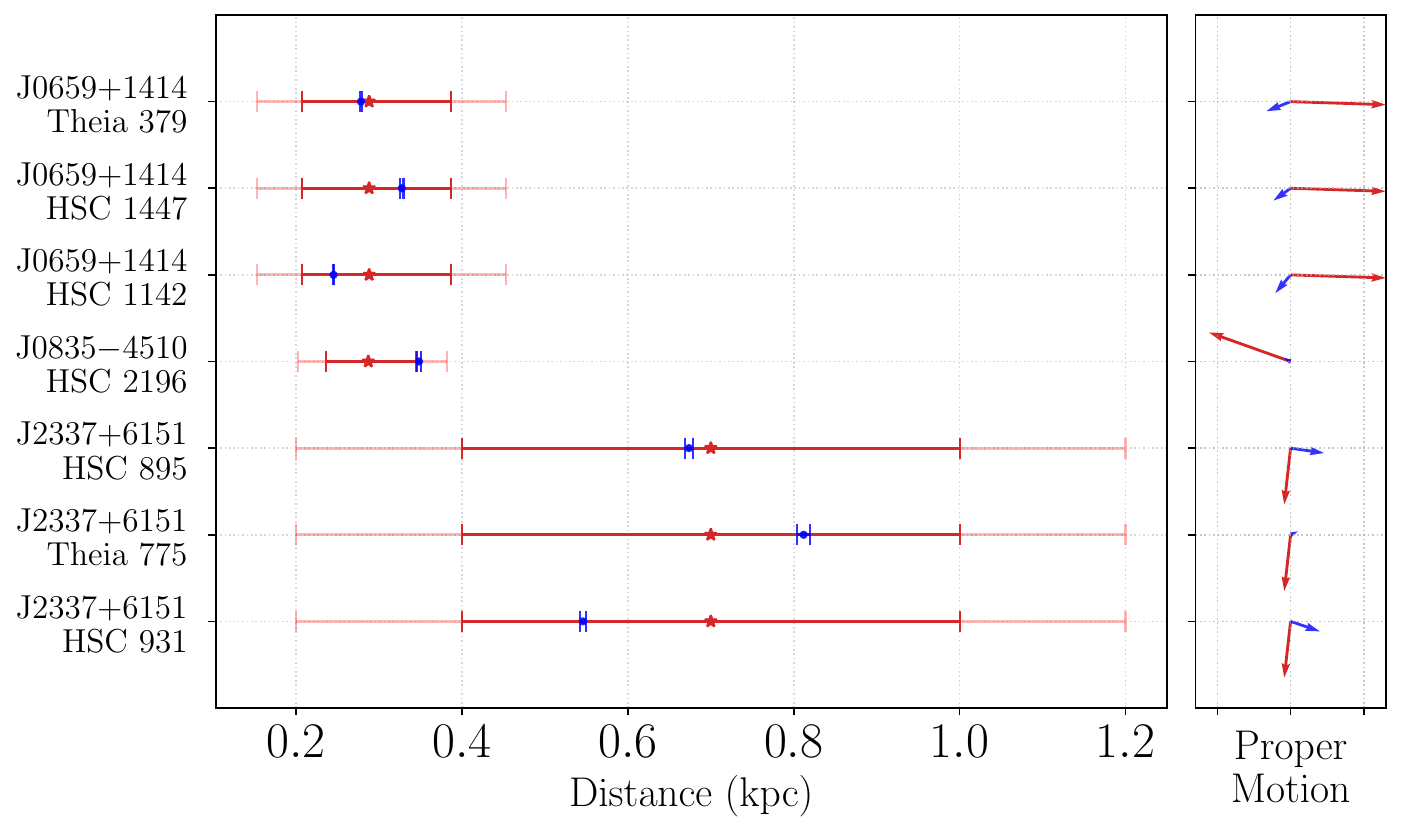}
    \caption{As is \autoref{fig:psr_oc_dist_pm} but for pulsar-moving group associations, showing (left) distance comparisons and (right) proper motion vector diagrams.}
    \label{fig:psr_mg_dist_pm}
\end{figure*}

\section{Potential Pulsar-Moving Group Associations at Present}
\label{append:psr-mg}

\begin{table*}[t]
    \centering
    \setlength{\tabcolsep}{6pt} 
    \renewcommand{\arraystretch}{1.1} 
    \begin{tabular}{ccccccccc}
        \hline\hline
        Name & Prob & $d$ & $\mu_{\alpha} \cos \delta$ & $\mu_{\delta}$ & Age & $P_0$ & $r_{50}$ & $M_{\rm tot}$ \\
         & ($\%$) & (kpc) & ($\operatorname{mas~yr}^{-1}$) & ($\operatorname{mas~yr}^{-1}$) & (yr) & (s) & (deg/pc) & ($\mathrm{M}_{\odot}$) \\
        \midrule 
        J0659+1414 & -- & $0.288^{+0.033}_{-0.027}$ & $-44.1 \pm 0.7$ & $-2.4 \pm 0.3$ & $1.11 \times 10^{5}$ & 0.38 & -- & -- \\ 
        Theia 379 & 99 (84) & $0.2781 \pm 0.0002$ & $-11.17 \pm 0.07$ & $-7.51 \pm 0.15$ & $1.40 \times 10^{8}$ & -- & 7.96/38.88 & 187.64 \\ 
        HSC 1447 & 98 (0) & $0.3274 \pm 0.0004$ & $-7.93 \pm 0.06$ & $-9.63 \pm 0.13$ & $1.37 \times 10^{8}$ & -- & 7.57/43.51 & 70.74 \\ 
        HSC 1142 & 83 (0) & $0.2451 \pm 0.0001$ & $-7.10 \pm 0.14$ & $-14.28 \pm 0.09$ & $4.31 \times 10^{8}$ & -- & 21.86/98.35 & 109.73 \\ 
        \midrule 
        J0835$-$4510 & \multirow{2}{*}{54 (4)} & $0.287^{+0.019}_{-0.017}$ & $-49.68 \pm 0.06$ & $29.90 \pm 0.10$ & $1.13 \times 10^{4}$ & 0.09 & -- & -- \\ 
        HSC 2196 & & $0.3481 \pm 0.0004$ & $-4.22 \pm 0.11$ & $3.54 \pm 0.05$ & $1.95 \times 10^{8}$ & -- & 4.00/24.33 &  57.45 \\ 
        \midrule 
        J2337+6151 & --  & $0.7 \pm 0.1$ & $-1 \pm 18$ & $-15 \pm 16$ & $4.06 \times 10^{4}$ & 0.50 & -- & --\\ 
        HSC 895 & 19 (0) & $0.6737 \pm 0.0009$ & $5.29 \pm 0.05$ & $-1.31 \pm 0.02$ & $1.70 \times 10^{8}$ & -- & 1.87/21.97 &206.84 \\ 
        Theia 775 & 17 (9) & $0.812 \pm 0.002$ &  $1.16 \pm 0.02$ & $0.89 \pm 0.02$ & $1.39 \times 10^{8}$ & -- & 1.06/15.05 & 181.54 \\ 
        HSC 931 & 8 (2) & $0.5460 \pm 0.0008$ & $4.61 \pm 0.03$ & $-2.67 \pm 0.06$ & $1.52 \times 10^{8}$ & -- & 1.43/13.59 & 119.01 \\ 
        \midrule 
    \end{tabular}
    \caption{As in ~\autoref{tab:current_psr_oc_pair}, but for pulsars and their possibly associated moving groups. The ``Prob" column shows the probability that a pulsar is within three times the moving group radius $3r_{50}$, with the values in parentheses corresponding to $2r_{50}$.}
    \label{tab:current_psr_mg_pair}
\end{table*}

By cross matching 164 pulsars and 539 moving groups, \autoref{tab:current_psr_mg_pair} lists seven candidate associations with $>1\%$ probabilities, and \autoref{fig:psr_mg_dist_pm} shows the distance estimates and proper motion characteristics of these pulsar-cluster pairs. Notably, two pairs show exceptionally high probabilities ($\geq 98\%$): PSR J0659+1414 with Theia~379 and HSC~1447.
The high association probability is a result of large angular sizes of moving groups; if we reduce the separation threshold from $3r_{50}$ to $2 r_{50}$, the overlap probability of J0659+1414 with HSC~1447 drops from $98\%$ to $0$.
Since moving groups are only candidate open clusters, we do not investigate pulsars listed in \autoref{tab:current_psr_mg_pair} further here.

\section{Additional candidate pulsar-open cluster associations in the past}
\label{append:psr-oc_past}

In \autoref{tab:past_psr_oc_pair_rv152}, we list 10 pulsar-open cluster pairs that have association probability $>15\%$ under the alternative pulsar radial velocity distribution model, i.e., a Gaussian distribution of $152 \pm 10 \, \mathrm{km \, s^{-1}}$ for normal pulsars or $54 \pm 6 \, \mathrm{km \, s^{-1}}$ for recycled pulsars.

\begin{table*}[t]
    \centering
    \setlength{\tabcolsep}{4.5pt}
    \renewcommand{\arraystretch}{1.1}
    \begin{tabular}{cccccccccc}
        \hline\hline
        Name & Prob & $d$ & $\mu_{\alpha} \cos \delta$ & $\mu_{\delta}$ & Age & $P_0$ & $r_{50}$ & $M_{\rm tot}$ \\
         & ($\%$) & (kpc) & ($\operatorname{mas~yr}^{-1}$) & ($\operatorname{mas~yr}^{-1}$) & (yr) & (s) & (deg/pc) & ($\mathrm{M}_{\odot}$) \\
        \midrule 
J0332$+$5434 & \multirow{2}{*}{30.6 (10.6)} & $1.68^{+0.07}_{-0.06}$ & $16.97 \pm 0.03$ & $-10.37 \pm 0.05$ & $5.53 \times 10^{6}$ & 0.71 & -- & -- \\
UBC~1577 & & $2.23 \pm 0.01$ & $-1.34 \pm 0.02$ & $-1.22 \pm 0.01$ & $1.13 \times 10^{8}$ & -- & 0.54/20.89 & 373.29 \\
        \midrule 
J1012$+$5307$^{\dagger}$ & -- & $0.877^{+0.035}_{-0.035}$ & $2.624 \pm 0.003$ & $-25.487 \pm 0.004$ & $4.86 \times 10^{9}$ & 0.01 & -- & -- \\
HSC~764 & 28.8 (5.1) & $0.865 \pm 0.001$ & $-2.44 \pm 0.01$ & $-4.46 \pm 0.04$ & $1.42 \times 10^{7}$ & -- & 3.05/46.18 & 334.76 \\
Theia~280 & 19.7 (3.5) & $0.814 \pm 0.001$ & $-1.66 \pm 0.03$ & $-5.31 \pm 0.03$ & $1.60 \times 10^{8}$ & -- & 2.41/18.67 & 702.66 \\
        \midrule 
J0335$+$4555 & \multirow{2}{*}{26.7 (9.8)} & $2.44^{+0.18}_{-0.12}$ & $-3.66 \pm 0.05$ & $-0.08 \pm 0.12$ & $5.80 \times 10^{8}$ & 0.27 & -- & -- \\
CWNU~2136 & & $1.2767 \pm 0.0056$ & $0.73 \pm 0.02$ & $-5.57 \pm 0.02$ & $3.31 \times 10^{8}$ & -- & 1.99/44.40 & 375.45 \\
        \midrule 
J0737$-$3039A$^{\dagger}$ & \multirow{2}{*}{21.4 (9.6)} & $1.1^{+0.2}_{-0.1}$ & $-3.8 \pm 0.7$ & $2.1 \pm 0.3$ & $2.04 \times 10^{8}$ & 0.02 & -- & -- \\
Theia~3397 & & $1.037 \pm 0.002$ & $-4.89 \pm 0.01$ & $4.01 \pm 0.01$ & $6.03 \times 10^{7}$ & -- & 1.17/21.15 & 482.84 \\
        \midrule 
J1713$+$0747$^{\dagger}$ & \multirow{2}{*}{21.1 (6.1)} & $1.05^{+0.06}_{-0.05}$ & $4.9215 \pm 0.0008$ & $-3.920 \pm 0.002$ & $8.49 \times 10^{9}$ & 0.005 & -- & -- \\
Theia~172 & & $0.4646 \pm 0.0004$ & $-6.46 \pm 0.02$ & $-5.04 \pm 0.02$ & $1.21 \times 10^{8}$ & -- & 1.09/8.82 & 396.42 \\
        \midrule 
J1856$-$3754 & \multirow{2}{*}{20.7 (12.8)} & $0.161^{+0.018}_{-0.014}$ & $325.9 \pm 0.3$ & $-59.22 \pm 0.18$ & $3.76 \times 10^{6}$ & 7.06 & -- & -- \\
OCSN~96 & & $0.14061 \pm 0.00004$ & $-11.54 \pm 0.09$ & $-23.89 \pm 0.06$ & $4.69 \times 10^{6}$ & -- & 1.62/3.99 & 132.30 \\
        \midrule
J1741$+$1351$^{\dagger}$ & \multirow{2}{*}{17.6 (2.9)} & $1.075^{+0.061}_{-0.055}$ & $-8.976 \pm 0.020$ & $-7.415 \pm 0.020$ & $1.96 \times 10^{9}$ & 0.004 & -- & -- \\
Theia~553 & & $0.705 \pm 0.001$ & $0.75 \pm 0.01$ & $-6.78 \pm 0.02$ & $9 \times 10^{7}$ & -- & 0.79/9.68 & 338.51 \\
        \midrule
J2113$+$2754 & \multirow{2}{*}{17.0 (6.2)} & $1.42 \pm 0.04$ & $-27.96 \pm 0.03$ & $-54.46 \pm 0.07$ & $7.27 \times 10^{6}$ & 1.20 & -- & -- \\
UBC~175 & & $1.298 \pm 0.002$ & $1.47 \pm 0.01$ & $-0.801 \pm 0.009$ & $3.59 \times 10^{8}$ & -- & 0.33/7.46 & 603.64 \\
        \midrule
J0636$+$5128$^{\dagger}$ & \multirow{2}{*}{15.1 (2.7)} & $0.204^{+0.029}_{-0.022}$ & $1.1 \pm 0.4$ & $-4.4 \pm 0.7$ & $1.32 \times 10^{10}$ & 0.003 & -- & -- \\
Theia~3475 & & $1.80 \pm 0.01$ & $0.15 \pm 0.01$ & $-2.00 \pm 0.01$ & $1.45 \times 10^{8}$ & -- & 0.25/7.96 & 379.62 \\
        \midrule
    \end{tabular}
    \caption{As \autoref{tab:past_psr_oc_pair} but for additional candidate pulsar-cluster association pairs identified from the alternative pulsar radial velocity distribution. In the association probability column, values outside parentheses are obtained from a Gaussian distribution of $152 \pm 10 \, \mathrm{km \, s^{-1}}$ for normal pulsars or $54 \pm 6 \, \mathrm{km \, s^{-1}}$ for recycled pulsars (names marked with the dagger symbol), while those in parentheses assumed a uniform distribution in the range of $[-200, 200]\,\mathrm{km\,s^{-1}}$.}
    \label{tab:past_psr_oc_pair_rv152}
\end{table*}

\clearpage

\bibliography{cite}{}

@ARTICLE{ZhangSB25,
       author = {{Zhang}, S.~B. and {Wei}, J.~J. and {Yang}, X. and {Dai}, S. and {Wang}, J.~S. and {Toomey}, L. and {Wang}, S.~Q. and {Hobbs}, G. and {Wu}, X.~F. and {Staveley-Smith}, L.},
        title = "{Searching for Radio Pulsars in Old Open Clusters from the Parkes Archive}",
      journal = {\apj},
         year = 2025,
        month = jul,
       volume = {988},
       number = {1},
          eid = {21},
        pages = {21},
          doi = {10.3847/1538-4357/ade802},
archivePrefix = {arXiv},
       eprint = {2506.19236},
 primaryClass = {astro-ph.HE},
       adsurl = {https://ui.adsabs.harvard.edu/abs/2025ApJ...988...21Z}
}

@ARTICLE{Fesen21AntliaSNR,
       author = {{Fesen}, Robert A. and {Drechsler}, Marcel and {Weil}, Kathryn E. and {Strottner}, Xavier and {Raymond}, John C. and {Rupert}, Justin and {Milisavljevic}, Dan and {Subrayan}, Bhagya M. and {di Cicco}, Dennis and {Walker}, Sean and {Mittelman}, David and {Ludgate}, Mathew},
        title = "{Far-UV and Optical Emissions from Three Very Large Supernova Remnants Located at Unusually High Galactic Latitudes}",
      journal = {\apj},
         year = 2021,
        month = oct,
       volume = {920},
       number = {2},
          eid = {90},
        pages = {90},
          doi = {10.3847/1538-4357/ac0ada},
archivePrefix = {arXiv},
       eprint = {2102.12599},
 primaryClass = {astro-ph.HE},
       adsurl = {https://ui.adsabs.harvard.edu/abs/2021ApJ...920...90F}
}

@ARTICLE{RBS+24,
       title = {The neutron star mass, distance, and inclination from precision timing of the brilliant millisecond pulsar j0437-4715},
      author = {Reardon, Daniel J and Bailes, Matthew and Shannon, Ryan M and Flynn, Chris and Askew, Jacob and Bhat, ND Ramesh and Chen, Zu-Cheng and Cury{\l}o, Ma{\l}gorzata and Feng, Yi and Hobbs, George B and others},
     journal = {The Astrophysical Journal Letters},
      volume = {971},
      number = {1},
       pages = {L18},
         doi = {10.3847/2041-8213/ad614a},
        year = {2024},
   publisher = {IOP Publishing},
      adsurl = {https://iopscience.iop.org/article/10.3847/2041-8213/ad614a}
}

@ARTICLE{JKC+18,
       title = {Binary pulsar distances and velocities from gaia data release 2},
      author = {Jennings, Ross J and Kaplan, David L and Chatterjee, Shami and Cordes, James M and Deller, Adam T},
     journal = {The Astrophysical Journal},
      volume = {864},
      number = {1},
       pages = {26},
         doi = {10.3847/1538-4357/aad084}, 
        year = {2018},
   publisher = {IOP Publishing},
      adsurl = {https://ui.adsabs.harvard.edu/link_gateway/2018ApJ...864...26J/doi:10.3847/1538-4357/aad084}
}

@ARTICLE{LSB+25,
       author = {{Liu}, Xiao-Jin and {Sengar}, Rahul and {Bailes}, Matthew and {Eatough}, Ralph P. and {Yuan}, Jianping and {Wang}, Na and {Zhu}, Weiwei and {Zhou}, Lu and {Gao}, He and {Zhu}, Zong-Hong and {Zhu}, Xing-Jiang},
        title = "{PSR J1922+37: a 1.9 s Pulsar Discovered in the Direction of the Old Open Cluster NGC 6791}",
      journal = {\apjl},
         year = 2025,
        month = mar,
       volume = {981},
       number = {2},
          eid = {L29},
        pages = {L29},
          doi = {10.3847/2041-8213/adb9c4},
archivePrefix = {arXiv},
       eprint = {2412.08055},
 primaryClass = {astro-ph.HE},
       adsurl = {https://ui.adsabs.harvard.edu/abs/2025ApJ...981L..29L}
}

@ARTICLE{VWC+12,
        title = {On Pulsar Distance Measurements and their Uncertainties},
       author = { Verbiest, J. P. W.  and  Weisberg, J. M.  and  Chael, A. A.  and  Lee, K. J.  and  Lorimer, D. R. },
      journal = {Astrophysical Journal},
       volume = {755},
       number = {1},
        pages = {-},
          doi = {10.1088/0004-637X/755/1/39},
         year = {2012},
       adsurl = {https://ui.adsabs.harvard.edu/abs/2012ApJ...755...39V/abstract}
}

@ARTICLE{HR24,
       author = {Hunt, Emily L and Reffert, Sabine},
        title = {Improving the open cluster census-III. Using cluster masses, radii, and dynamics to create a cleaned open cluster catalogue},
      journal = {Astronomy \& Astrophysics},
       volume = {686},
        pages = {A42},
          doi = {https://doi.org/10.1051/0004-6361/202348662},
         year = 2024,
       adsurl = {https://ui.adsabs.harvard.edu/abs/2024A%26A...686A..42H/abstract}
}

@ARTICLE{YMW17,
       author = {{Yao}, J.~M. and {Manchester}, R.~N. and {Wang}, N.},
        title = "{A New Electron-density Model for Estimation of Pulsar and FRB Distances}",
      journal = {\apj},
         year = 2017,
        month = jan,
       volume = {835},
       number = {1},
          eid = {29},
        pages = {29},
          doi = {10.3847/1538-4357/835/1/29},
       eprint = {1610.09448},
 primaryClass = {astro-ph.GA},
       adsurl = {https://ui.adsabs.harvard.edu/abs/2017ApJ...835...29Y}
}

@ARTICLE{SBF+24,
        title = {MeerKAT Pulsar Timing Array parallaxes and proper motions},
       author = {Shamohammadi, Mohsen and Bailes, Matthew and Flynn, Christopher and Reardon, Daniel J and Shannon, Ryan M and Buchner, Sarah and Cameron, Andrew D and Camilo, Fernando and Coronigu, Alessandro and Geyer, Marisa and others},
      journal = {Monthly Notices of the Royal Astronomical Society},
       volume = {530},
       number = {1},
        pages = {287--306},
          doi = {10.1093/mnras/stae016},
         year = {2024},
    publisher = {Oxford University Press},
       adsurl = {https://doi.org/10.1093/mnras/stae2578}
}

@ARTICLE{NBB+14,
        title = {The High Time Resolution Universe pulsar survey--X. Discovery of four millisecond pulsars and updated timing solutions of a further 12},
       author = {Ng, C and Bailes, M and Bates, SD and Bhat, NDR and Burgay, MARTA and Burke-Spolaor, S and Champion, DJ and Coster, P and Johnston, S and Keith, MJ and others},
      journal = {Monthly Notices of the Royal Astronomical Society},
       volume = {439},
       number = {2},
        pages = {1865--1883},
          doi = {10.1093/mnras/stu067},
         year = {2014},
    publisher = {The Royal Astronomical Society},
       adsurl = {https://academic.oup.com/mnras/article/439/2/1865/1014482}
}

@ARTICLE{VBC+09,
        title = {Timing stability of millisecond pulsars and prospects for gravitational-wave detection},
       author = {Verbiest, JPW and Bailes, M and Coles, WA and Hobbs, GB and Van Straten, W and Champion, DJ and Jenet, FA and Manchester, RN and Bhat, NDR and Sarkissian, JM and others},
      journal = {Monthly Notices of the Royal Astronomical Society},
       volume = {400},
       number = {2},
        pages = {951--968},
          doi = {10.1111/j.1365-2966.2009.15508.x},
         year = {2009},
    publisher = {Blackwell Publishing Ltd Oxford, UK},
       adsurl = {https://ui.adsabs.harvard.edu/abs/2009MNRAS.400..951V/abstract}
}

@ARTICLE{MHT+05,
       author = {{Manchester}, R.~N. and {Hobbs}, G.~B. and {Teoh}, A. and {Hobbs}, M.},
        title = "{The Australia Telescope National Facility Pulsar Catalogue}",
      journal = {\aj},
         year = 2005,
        month = apr,
       volume = {129},
       number = {4},
        pages = {1993-2006},
          doi = {10.1086/428488},
archivePrefix = {arXiv},
       eprint = {astro-ph/0412641},
 primaryClass = {astro-ph},
       adsurl = {https://ui.adsabs.harvard.edu/abs/2005AJ....129.1993M}
}

@ARTICLE{Hobbs2005,
       author = {{Hobbs}, G. and {Lorimer}, D.~R. and {Lyne}, A.~G. and {Kramer}, M.},
        title = "{A statistical study of 233 pulsar proper motions}",
      journal = {\mnras},
         year = 2005,
        month = jul,
       volume = {360},
       number = {3},
        pages = {974-992},
          doi = {10.1111/j.1365-2966.2005.09087.x},
archivePrefix = {arXiv},
       eprint = {astro-ph/0504584},
 primaryClass = {astro-ph},
       adsurl = {https://ui.adsabs.harvard.edu/abs/2005MNRAS.360..974H}
}

@ARTICLE{Astropy13,
       author = {{Astropy Collaboration} and {Robitaille}, Thomas P. and {Tollerud}, Erik J. and {Greenfield}, Perry and {Droettboom}, Michael and {Bray}, Erik and {Aldcroft}, Tom and {Davis}, Matt and {Ginsburg}, Adam and {Price-Whelan}, Adrian M. and {Kerzendorf}, Wolfgang E. and {Conley}, Alexander and {Crighton}, Neil and {Barbary}, Kyle and {Muna}, Demitri and {Ferguson}, Henry and {Grollier}, Fr{\'e}d{\'e}ric and {Parikh}, Madhura M. and {Nair}, Prasanth H. and {Unther}, Hans M. and {Deil}, Christoph and {Woillez}, Julien and {Conseil}, Simon and {Kramer}, Roban and {Turner}, James E.~H. and {Singer}, Leo and {Fox}, Ryan and {Weaver}, Benjamin A. and {Zabalza}, Victor and {Edwards}, Zachary I. and {Azalee Bostroem}, K. and {Burke}, D.~J. and {Casey}, Andrew R. and {Crawford}, Steven M. and {Dencheva}, Nadia and {Ely}, Justin and {Jenness}, Tim and {Labrie}, Kathleen and {Lim}, Pey Lian and {Pierfederici}, Francesco and {Pontzen}, Andrew and {Ptak}, Andy and {Refsdal}, Brian and {Servillat}, Mathieu and {Streicher}, Ole},
        title = "{Astropy: A community Python package for astronomy}",
      journal = {\aap},
         year = 2013,
        month = oct,
       volume = {558},
          eid = {A33},
        pages = {A33},
          doi = {10.1051/0004-6361/201322068},
archivePrefix = {arXiv},
       eprint = {1307.6212},
 primaryClass = {astro-ph.IM},
       adsurl = {https://ui.adsabs.harvard.edu/abs/2013A&A...558A..33A}
}

@ARTICLE{Astropy18,
       author = {{Astropy Collaboration} and {Price-Whelan}, A.~M. and {Sip{\H{o}}cz}, B.~M. and {G{\"u}nther}, H.~M. and {Lim}, P.~L. and {Crawford}, S.~M. and {Conseil}, S. and {Shupe}, D.~L. and {Craig}, M.~W. and {Dencheva}, N. and {Ginsburg}, A. and {VanderPlas}, J.~T. and {Bradley}, L.~D. and {P{\'e}rez-Su{\'a}rez}, D. and {de Val-Borro}, M. and {Aldcroft}, T.~L. and {Cruz}, K.~L. and {Robitaille}, T.~P. and {Tollerud}, E.~J. and {Ardelean}, C. and {Babej}, T. and {Bach}, Y.~P. and {Bachetti}, M. and {Bakanov}, A.~V. and {Bamford}, S.~P. and {Barentsen}, G. and {Barmby}, P. and {Baumbach}, A. and {Berry}, K.~L. and {Biscani}, F. and {Boquien}, M. and {Bostroem}, K.~A. and {Bouma}, L.~G. and {Brammer}, G.~B. and {Bray}, E.~M. and {Breytenbach}, H. and {Buddelmeijer}, H. and {Burke}, D.~J. and {Calderone}, G. and {Cano Rodr{\'\i}guez}, J.~L. and {Cara}, M. and {Cardoso}, J.~V.~M. and {Cheedella}, S. and {Copin}, Y. and {Corrales}, L. and {Crichton}, D. and {D'Avella}, D. and {Deil}, C. and {Depagne}, {\'E}. and {Dietrich}, J.~P. and {Donath}, A. and {Droettboom}, M. and {Earl}, N. and {Erben}, T. and {Fabbro}, S. and {Ferreira}, L.~A. and {Finethy}, T. and {Fox}, R.~T. and {Garrison}, L.~H. and {Gibbons}, S.~L.~J. and {Goldstein}, D.~A. and {Gommers}, R. and {Greco}, J.~P. and {Greenfield}, P. and {Groener}, A.~M. and {Grollier}, F. and {Hagen}, A. and {Hirst}, P. and {Homeier}, D. and {Horton}, A.~J. and {Hosseinzadeh}, G. and {Hu}, L. and {Hunkeler}, J.~S. and {Ivezi{\'c}}, {\v{Z}}. and {Jain}, A. and {Jenness}, T. and {Kanarek}, G. and {Kendrew}, S. and {Kern}, N.~S. and {Kerzendorf}, W.~E. and {Khvalko}, A. and {King}, J. and {Kirkby}, D. and {Kulkarni}, A.~M. and {Kumar}, A. and {Lee}, A. and {Lenz}, D. and {Littlefair}, S.~P. and {Ma}, Z. and {Macleod}, D.~M. and {Mastropietro}, M. and {McCully}, C. and {Montagnac}, S. and {Morris}, B.~M. and {Mueller}, M. and {Mumford}, S.~J. and {Muna}, D. and {Murphy}, N.~A. and {Nelson}, S. and {Nguyen}, G.~H. and {Ninan}, J.~P. and {N{\"o}the}, M. and {Ogaz}, S. and {Oh}, S. and {Parejko}, J.~K. and {Parley}, N. and {Pascual}, S. and {Patil}, R. and {Patil}, A.~A. and {Plunkett}, A.~L. and {Prochaska}, J.~X. and {Rastogi}, T. and {Reddy Janga}, V. and {Sabater}, J. and {Sakurikar}, P. and {Seifert}, M. and {Sherbert}, L.~E. and {Sherwood-Taylor}, H. and {Shih}, A.~Y. and {Sick}, J. and {Silbiger}, M.~T. and {Singanamalla}, S. and {Singer}, L.~P. and {Sladen}, P.~H. and {Sooley}, K.~A. and {Sornarajah}, S. and {Streicher}, O. and {Teuben}, P. and {Thomas}, S.~W. and {Tremblay}, G.~R. and {Turner}, J.~E.~H. and {Terr{\'o}n}, V. and {van Kerkwijk}, M.~H. and {de la Vega}, A. and {Watkins}, L.~L. and {Weaver}, B.~A. and {Whitmore}, J.~B. and {Woillez}, J. and {Zabalza}, V. and {Astropy Contributors}},
        title = "{The Astropy Project: Building an Open-science Project and Status of the v2.0 Core Package}",
      journal = {\aj},
         year = 2018,
        month = sep,
       volume = {156},
       number = {3},
          eid = {123},
        pages = {123},
          doi = {10.3847/1538-3881/aabc4f},
archivePrefix = {arXiv},
       eprint = {1801.02634},
 primaryClass = {astro-ph.IM},
       adsurl = {https://ui.adsabs.harvard.edu/abs/2018AJ....156..123A}
}

@ARTICLE{Astropy22,
       author = {{Astropy Collaboration} and {Price-Whelan}, Adrian M. and {Lim}, Pey Lian and {Earl}, Nicholas and {Starkman}, Nathaniel and {Bradley}, Larry and {Shupe}, David L. and {Patil}, Aarya A. and {Corrales}, Lia and {Brasseur}, C.~E. and {N{\"o}the}, Maximilian and {Donath}, Axel and {Tollerud}, Erik and {Morris}, Brett M. and {Ginsburg}, Adam and {Vaher}, Eero and {Weaver}, Benjamin A. and {Tocknell}, James and {Jamieson}, William and {van Kerkwijk}, Marten H. and {Robitaille}, Thomas P. and {Merry}, Bruce and {Bachetti}, Matteo and {G{\"u}nther}, H. Moritz and {Aldcroft}, Thomas L. and {Alvarado-Montes}, Jaime A. and {Archibald}, Anne M. and {B{\'o}di}, Attila and {Bapat}, Shreyas and {Barentsen}, Geert and {Baz{\'a}n}, Juanjo and {Biswas}, Manish and {Boquien}, M{\'e}d{\'e}ric and {Burke}, D.~J. and {Cara}, Daria and {Cara}, Mihai and {Conroy}, Kyle E. and {Conseil}, Simon and {Craig}, Matthew W. and {Cross}, Robert M. and {Cruz}, Kelle L. and {D'Eugenio}, Francesco and {Dencheva}, Nadia and {Devillepoix}, Hadrien A.~R. and {Dietrich}, J{\"o}rg P. and {Eigenbrot}, Arthur Davis and {Erben}, Thomas and {Ferreira}, Leonardo and {Foreman-Mackey}, Daniel and {Fox}, Ryan and {Freij}, Nabil and {Garg}, Suyog and {Geda}, Robel and {Glattly}, Lauren and {Gondhalekar}, Yash and {Gordon}, Karl D. and {Grant}, David and {Greenfield}, Perry and {Groener}, Austen M. and {Guest}, Steve and {Gurovich}, Sebastian and {Handberg}, Rasmus and {Hart}, Akeem and {Hatfield-Dodds}, Zac and {Homeier}, Derek and {Hosseinzadeh}, Griffin and {Jenness}, Tim and {Jones}, Craig K. and {Joseph}, Prajwel and {Kalmbach}, J. Bryce and {Karamehmetoglu}, Emir and {Ka{\l}uszy{\'n}ski}, Miko{\l}aj and {Kelley}, Michael S.~P. and {Kern}, Nicholas and {Kerzendorf}, Wolfgang E. and {Koch}, Eric W. and {Kulumani}, Shankar and {Lee}, Antony and {Ly}, Chun and {Ma}, Zhiyuan and {MacBride}, Conor and {Maljaars}, Jakob M. and {Muna}, Demitri and {Murphy}, N.~A. and {Norman}, Henrik and {O'Steen}, Richard and {Oman}, Kyle A. and {Pacifici}, Camilla and {Pascual}, Sergio and {Pascual-Granado}, J. and {Patil}, Rohit R. and {Perren}, Gabriel I. and {Pickering}, Timothy E. and {Rastogi}, Tanuj and {Roulston}, Benjamin R. and {Ryan}, Daniel F. and {Rykoff}, Eli S. and {Sabater}, Jose and {Sakurikar}, Parikshit and {Salgado}, Jes{\'u}s and {Sanghi}, Aniket and {Saunders}, Nicholas and {Savchenko}, Volodymyr and {Schwardt}, Ludwig and {Seifert-Eckert}, Michael and {Shih}, Albert Y. and {Jain}, Anany Shrey and {Shukla}, Gyanendra and {Sick}, Jonathan and {Simpson}, Chris and {Singanamalla}, Sudheesh and {Singer}, Leo P. and {Singhal}, Jaladh and {Sinha}, Manodeep and {Sip{\H{o}}cz}, Brigitta M. and {Spitler}, Lee R. and {Stansby}, David and {Streicher}, Ole and {{\v{S}}umak}, Jani and {Swinbank}, John D. and {Taranu}, Dan S. and {Tewary}, Nikita and {Tremblay}, Grant R. and {de Val-Borro}, Miguel and {Van Kooten}, Samuel J. and {Vasovi{\'c}}, Zlatan and {Verma}, Shresth and {de Miranda Cardoso}, Jos{\'e} Vin{\'\i}cius and {Williams}, Peter K.~G. and {Wilson}, Tom J. and {Winkel}, Benjamin and {Wood-Vasey}, W.~M. and {Xue}, Rui and {Yoachim}, Peter and {Zhang}, Chen and {Zonca}, Andrea and {Astropy Project Contributors}},
        title = "{The Astropy Project: Sustaining and Growing a Community-oriented Open-source Project and the Latest Major Release (v5.0) of the Core Package}",
      journal = {\apj},
         year = 2022,
        month = aug,
       volume = {935},
       number = {2},
          eid = {167},
        pages = {167},
          doi = {10.3847/1538-4357/ac7c74},
archivePrefix = {arXiv},
       eprint = {2206.14220},
 primaryClass = {astro-ph.IM},
       adsurl = {https://ui.adsabs.harvard.edu/abs/2022ApJ...935..167A}
}

@ARTICLE{Hunter07,
       author = {{Hunter}, John D.},
        title = "{Matplotlib: A 2D Graphics Environment}",
      journal = {Computing in Science and Engineering},
         year = 2007,
        month = may,
       volume = {9},
       number = {3},
        pages = {90-95},
          doi = {10.1109/MCSE.2007.55},
       adsurl = {https://ui.adsabs.harvard.edu/abs/2007CSE.....9...90H}
}

@ARTICLE{HMv+20,
       author = {{Harris}, Charles R. and {Millman}, K. Jarrod and {van der Walt}, St{\'e}fan J. and {Gommers}, Ralf and {Virtanen}, Pauli and {Cournapeau}, David and {Wieser}, Eric and {Taylor}, Julian and {Berg}, Sebastian and {Smith}, Nathaniel J. and {Kern}, Robert and {Picus}, Matti and {Hoyer}, Stephan and {van Kerkwijk}, Marten H. and {Brett}, Matthew and {Haldane}, Allan and {del R{\'\i}o}, Jaime Fern{\'a}ndez and {Wiebe}, Mark and {Peterson}, Pearu and {G{\'e}rard-Marchant}, Pierre and {Sheppard}, Kevin and {Reddy}, Tyler and {Weckesser}, Warren and {Abbasi}, Hameer and {Gohlke}, Christoph and {Oliphant}, Travis E.},
        title = "{Array programming with NumPy}",
      journal = {\nat},
         year = 2020,
        month = sep,
       volume = {585},
       number = {7825},
        pages = {357-362},
          doi = {10.1038/s41586-020-2649-2},
archivePrefix = {arXiv},
       eprint = {2006.10256},
 primaryClass = {cs.MS},
       adsurl = {https://ui.adsabs.harvard.edu/abs/2020Natur.585..357H}
}

@ARTICLE{DP18,
        title = {On the solar velocity},
       author = {Drimmel, Ronald and Poggio, Eloisa},
      journal = {AAS},
       volume = {2},
       number = {4},
        pages = {210},
          doi = {10.3847/2515-5172/aaef8b},
         year = {2018},
       adsurl = {https://iopscience.iop.org/article/10.3847/2515-5172/aaef8b}
}

@ARTICLE{2019_solar,
        title = {Vertical waves in the solar neighbourhood in Gaia DR2},
       author = {Bennett, Morgan and Bovy, Jo},
      journal = {Monthly Notices of the Royal Astronomical Society},
       volume = {482},
       number = {1},
        pages = {1417--1425},
          doi = {10.1093/mnras/sty2813},
         year = {2019},
    publisher = {Oxford University Press},
       adsurl = {https://academic.oup.com/mnras/article/482/1/1417/5142313?login=false}
}

@ARTICLE{DDS+23,
        title = {The MSPSR$\pi$ catalogue: VLBA astrometry of 18 millisecond pulsars},
       author = {Ding, H and Deller, AT and Stappers, BW and Lazio, TJW and Kaplan, D and Chatterjee, S and Brisken, W and Cordes, J and Freire, PCC and Fonseca, E and others},
      journal = {Monthly Notices of the Royal Astronomical Society},
       volume = {519},
       number = {4},
        pages = {4982--5007},
          doi = {10.1093/mnras/stac3725},
         year = {2023},
    publisher = {Oxford University Press},
       adsurl = {https://academic.oup.com/mnras/article/519/4/4982/6948353?login=false}
}

@ARTICLE{DGB+19,
        title = {Microarcsecond VLBI pulsar astrometry with PSR$\pi$ II. Parallax distances for 57 pulsars},
       author = {Deller, AT and Goss, WM and Brisken, WF and Chatterjee, S and Cordes, JM and Janssen, GH and Kovalev, YY and Lazio, TJW and Petrov, L and Stappers, BW and others},
      journal = {The Astrophysical Journal},
       volume = {875},
       number = {2},
        pages = {100},
          doi = {10.3847/1538-4357/ab11c7},
         year = {2019},
    publisher = {IOP Publishing},
       adsurl = {https://iopscience.iop.org/article/10.3847/1538-4357/ab11c7}
}

@ARTICLE{BTG+03,
        title = {The distance and radius of the neutron star PSR B0656+ 14},
       author = {Brisken, Walter F and Thorsett, SE and Golden, A and Goss, W Miller},
      journal = {The Astrophysical Journal},
       volume = {593},
       number = {2},
        pages = {L89},
          doi = {10.1086/378184},
         year = {2003},
    publisher = {IOP Publishing},
       adsurl = {https://ui.adsabs.harvard.edu/abs/2003ApJ...593L..89B/abstract}
}

@ARTICLE{BBG+02,
        title = {Very long baseline array measurement of nine pulsar parallaxes},
       author = {Brisken, Walter F and Benson, John M and Goss, W Miller and Thorsett, SE},
      journal = {The Astrophysical Journal},
       volume = {571},
       number = {2},
        pages = {906},
          doi = {10.1086/340098},
         year = {2002},
    publisher = {IOP Publishing},
       adsurl = {https://ui.adsabs.harvard.edu/abs/2002ApJ...571..906B/abstract}
}

@ARTICLE{DLR+03,
        title = {The Vela pulsar’s proper motion and parallax derived from VLBI observations},
       author = {Dodson, Richard and Legge, D and Reynolds, JE and McCulloch, Peter M},
      journal = {The Astrophysical Journal},
       volume = {596},
       number = {2},
        pages = {1137},
          doi = {10.1086/378089},
         year = {2003},
    publisher = {IOP Publishing},
       adsurl = {https://ui.adsabs.harvard.edu/abs/2003ApJ...596.1137D/abstract}
}

@ARTICLE{LPM+16,
        title = {Pulsar lensing geometry},
       author = {Liu, Siqi and Pen, Ue-Li and Macquart, J-P and Brisken, Walter and Deller, Adam},
      journal = {Monthly Notices of the Royal Astronomical Society},
       volume = {458},
       number = {2},
        pages = {1289--1299},
          doi = {10.1093/mnras/stw314},
         year = {2016},
    publisher = {The Royal Astronomical Society},
       adsurl = {https://ui.adsabs.harvard.edu/abs/2016MNRAS.458.1289L/abstract}
}

@ARTICLE{GSL+16,
        title = {The gamma-ray millisecond pulsar deathline, revisited-New velocity and distance measurements},
       author = {Guillemot, Lucas and Smith, DA and Laffon, H and Janssen, GH and Cognard, Isma{\"e}l and Theureau, Gilles and Desvignes, Gr{\'e}gory and Ferrara, EC and Ray, PS},
      journal = {Astronomy \& Astrophysics},
       volume = {587},
        pages = {A109},
          doi = {10.1051/0004-6361/201527847},
         year = {2016},
    publisher = {EDP Sciences},
       adsurl = {https://www.aanda.org/articles/aa/full_html/2016/03/aa27847-15/aa27847-15.html}
}

@ARTICLE{VK07,
       author = {{van Kerkwijk}, M.~H. and {Kaplan}, D.~L.},
        title = "{Isolated neutron stars: magnetic fields, distances,and spectra}",
      journal = {\apss},
         year = 2007,
        month = apr,
       volume = {308},
       number = {1-4},
        pages = {191-201},
          doi = {10.1007/s10509-007-9343-9},
archivePrefix = {arXiv},
       eprint = {astro-ph/0607320},
 primaryClass = {astro-ph},
       adsurl = {https://ui.adsabs.harvard.edu/abs/2007Ap&SS.308..191V}
}

@ARTICLE{HBZ01,
       author = {Hoogerwerf, R and De Bruijne, JHJ and De Zeeuw, PT},
        title = {On the origin of the O and B-type stars with high velocities-II. Runaway stars and pulsars ejected from the nearby young stellar groups},
      journal = {Astronomy \& Astrophysics},
       volume = {365},
       number = {2},
        pages = {49-77},
          doi = {10.1051/0004-6361:20000014},
         year = 2001,
       adsurl = {https://www.aanda.org/articles/aa/abs/2001/02/aa10198/aa10198.html}
}

@ARTICLE{THS+12,
       author = {Tetzlaff, Nina and Schmidt, Janos G and Hohle, Markus M and Neuhaeuser, Ralph},
        title = {Neutron stars from young nearby associations: The origin of RX J1605. 3+ 3249},
      journal = {Publications of the Astronomical Society of Australia},
       volume = {29},
       number = {2},
        pages = {98-108},
          doi = {10.1071/AS11057},
         year = 2012,
       adsurl = {https://ui.adsabs.harvard.edu/abs/2012PASA...29...98T/abstract}
}

@ARTICLE{TTN+13,
       author = {Tetzlaff, Nina and Torres, Guillermo and Neuhaeuser, Ralph and Hohle, Markus Matthias},
        title = {The neutron star born in the Antlia supernova remnant},
      journal = {Monthly Notices of the Royal Astronomical Society},
       volume = {435},
       number = {1},
        pages = {879--884},
          doi = {10.1093/mnras/stt1358},
         year = 2013,
       adsurl = {https://academic.oup.com/mnras/article/435/1/879/1131044}
}

@ARTICLE{TPC+11,
       title = {PSRs J0248+ 6021 and J2240+ 5832: young pulsars in the northern Galactic plane-Discovery, timing, and gamma-ray observations},
      author = {Theureau, Gilles and Parent, D and Cognard, Isma{\"e}l and Desvignes, G and Smith, DA and Casandjian, JM and Cheung, CC and Craig, HA and Donato, D and Foster, R and others},
     journal = {Astronomy \& Astrophysics},
      volume = {525},
       pages = {A94},
         doi = {10.1051/0004-6361/201015317},
        year = {2011},
   publisher = {EDP Sciences},
      adsurl = {https://www.aanda.org/articles/aa/full_html/2011/01/aa15317-10/aa15317-10.html}
}

@ARTICLE{CCL+69,
       title = {Crab nebula pulsar NP 0532},
      author = {Comella, JM and Craft, HD and Lovelace, RVE and Sutton, JM and Tyler, G Leonard},
     journal = {Nature},
      volume = {221},
      number = {5179},
       pages = {453--454},
         doi = {10.1038/221453a0},
        year = {1969},
   publisher = {Nature Publishing Group UK London},
      adsurl = {https://ui.adsabs.harvard.edu/abs/1969Natur.221..453C/abstrac}
}

@ARTICLE{LVM68,
       title = {A pulsar Suprenova Association?},
      author = {Large, MI and Vaughan, AE and MILLS, BY},
     journal = {Nature},
      volume = {220},
       pages = {340--341},
         doi = {10.1038/220340a0},
        year = {1968},
   publisher = {Nature Publishing Group UK London},
      adsurl = {https://www.nature.com/articles/220340a0#citeas}
}

@ARTICLE{MDS+18,
       title = {The geometric distance and binary orbit of PSR B1259--63},
      author = {Miller-Jones, James CA and Deller, Adam T and Shannon, Ryan M and Dodson,  Richard and Mold{\'o}n, Javier and Rib{\'o}, Marc and Dubus, Guillaume and Johnston, Simon and Paredes, Josep M and Ransom, Scott M and others},
     journal = {Monthly Notices of the Royal Astronomical Society},
      volume = {479},
      number = {4},
       pages = {4849--4860},
         doi = {10.1093/mnras/sty1775},
        year = {2018},
   publisher = {Oxford University Press},
      adsurl = {https://academic.oup.com/mnras/article/479/4/4849/5049310?login=false}
}

@ARTICLE{JML+92,
       title = {PSR 1259-63 : a binary radio pulsar with a Be star companion},
      author = {Simon Johnston and R. N. Manchester and Andrew G. Lyne and Matthew Bailes and Victoria M. Kaspi and Guo-jun Qiao and Nicolo' D'Amico},
     journal = {The Astrophysical Journal},
        year = {1992},
      volume = {387},
       pages = {L37},
         doi = {10.1086/186300},
       asurl = {https://api.semanticscholar.org/CorpusID:120184393}
}

@ARTICLE{JML+94,
      author = {{Johnston}, S. and {Manchester}, R.~N. and {Lyne}, A.~G. and {Nicastro}, L. and {Spyromilio}, J.},
       title = "{Radio and Optical Observations of the PSR:B1259-63 / SS:2883 Be-Star Binary System}",
     journal = {\mnras},
        year = 1994,
      volume = {268},
       pages = {430},
         doi = {10.1093/mnras/268.2.430},
      adsurl = {https://ui.adsabs.harvard.edu/abs/1994MNRAS.268..430J}
}

@ARTICLE{NRH+11,
       title = {Astrophysical parameters of LS 2883 and implications for the PSR B1259--63 gamma-ray binary},
      author = {Negueruela, Ignacio and Rib{\'o}, Marc and Herrero, Artemio and Lorenzo, Javier and Khangulyan, Dmitry and Aharonian, Felix A},
     journal = {The Astrophysical Journal Letters},
      volume = {732},
      number = {1},
       pages = {L11},
         doi = {10.1088/2041-8205/732/1/L11},
        year = {2011},
   publisher = {IOP Publishing},
      adsurl = {https://ui.adsabs.harvard.edu/abs/2011ApJ...732L..11N}
}

@ARTICLE{SGM14,
       title = {The kinematics and orbital dynamics of the PSR B1259- 63/LS 2883 system from 23 yr of pulsar timing},
      author = {Shannon, RM and Johnston, S and Manchester, RN},
     journal = {Monthly Notices of the Royal Astronomical Society},
      volume = {437},
      number = {4},
       pages = {3255--3264},
         doi = {10.1093/mnras/stt2123},
        year = {2014},
   publisher = {The Royal Astronomical Society},
      adsurl = {https://academic.oup.com/mnras/article/437/4/3255/1001679}
}

@ARTICLE{CO13,
       title = {Members of Centaurus OB1 and NGC 4755: new spectroscopic and astrometric studies},
      author = {Corti, Mariela Alejandra and Orellana, Rosa Beatriz},
     journal = {Astronomy \& Astrophysics},
      volume = {553},
       pages = {A108},
         doi = {10.1051/0004-6361/201220743},
        year = {2013},
   publisher = {Edp Sciences},
      adsurl = {https://www.aanda.org/articles/aa/full_html/2013/05/aa20743-12/aa20743-12.html}
}

@ARTICLE{LS22,
       author = {{Li}, Lu and {Shao}, Zhengyi},
        title = "{MiMO: Mixture Model for Open Clusters in Color-Magnitude Diagrams}",
      journal = {\apj},
         year = 2022,
        month = may,
       volume = {930},
       number = {1},
          eid = {44},
        pages = {44},
          doi = {10.3847/1538-4357/ac5f4f},
archivePrefix = {arXiv},
       eprint = {2112.08028},
 primaryClass = {astro-ph.GA},
       adsurl = {https://ui.adsabs.harvard.edu/abs/2022ApJ...930...44L}
}

@ARTICLE{K62,
       author = {{King}, Ivan},
        title = "{The structure of star clusters. I. an empirical density law}",
      journal = {\aj},
         year = 1962,
        month = oct,
       volume = {67},
        pages = {471},
          doi = {10.1086/108756},
       adsurl = {https://ui.adsabs.harvard.edu/abs/1962AJ.....67..471K}
}

@ARTICLE{JHB96,
       author = {{Johnston}, Kathryn V. and {Hernquist}, Lars and {Bolte}, Michael},
        title = "{Fossil Signatures of Ancient Accretion Events in the Halo}",
      journal = {\apj},
         year = 1996,
        month = jul,
       volume = {465},
        pages = {278},
          doi = {10.1086/177418},
archivePrefix = {arXiv},
       eprint = {astro-ph/9602060},
 primaryClass = {astro-ph},
       adsurl = {https://ui.adsabs.harvard.edu/abs/1996ApJ...465..278J}
}

@ARTICLE{P17,
       author = {{Price-Whelan}, Adrian M.},
        title = "{Gala: A Python package for galactic dynamics}",
      journal = {The Journal of Open Source Software},
         year = 2017,
        month = oct,
       volume = {2},
          eid = {388},
        pages = {388},
          doi = {10.21105/joss.00388},
       adsurl = {https://ui.adsabs.harvard.edu/abs/2017JOSS....2..388P}
}

@ARTICLE{B15,
        title = {galpy: A python Library for Galactic Dynamics},
       author = {Bovy, Jo},
      journal = {The Astrophysical Journal Supplement Series},
       volume = {216},
       number = {2},
        pages = {29},
          doi = {10.1088/0067-0049/216/2/29},
         year = {2015},
    publisher = {IOP Publishing},
       adsurl = {https://iopscience.iop.org/article/10.1088/0067-0049/216/2/29/meta}
}

@ARTICLE{FRS93,
       author = {{Furst}, E. and {Reich}, W. and {Seiradakis}, J.~H.},
        title = "{A new pulsar - supernova remnant association : PSR 2334+61 and G 114.3+0.3.}",
      journal = {\aap},
         year = 1993,
        month = sep,
       volume = {276},
        pages = {470-472},
       adsurl = {https://ui.adsabs.harvard.edu/abs/1993A&A...276..470F}
}

@ARTICLE{KPH+93,
       author = {{Kulkarni}, S.~R. and {Predehl}, P. and {Hasinger}, G. and {Aschenbach}, B.},
        title = "{An association between a long-period pulsar and an old supernova remnant}",
      journal = {\nat},
         year = 1993,
        month = mar,
       volume = {362},
       number = {6416},
        pages = {135-137},
          doi = {10.1038/362135a0},
       adsurl = {https://ui.adsabs.harvard.edu/abs/1993Natur.362..135K}
}

@ARTICLE{Cavallo2024,
       author = {{Cavallo}, Lorenzo and {Spina}, Lorenzo and {Carraro}, Giovanni and {Magrini}, Laura and {Poggio}, Eloisa and {Cantat-Gaudin}, Tristan and {Pasquato}, Mario and {Lucatello}, Sara and {Ortolani}, Sergio and {Schiappacasse-Ulloa}, Jose},
        title = "{Parameter Estimation for Open Clusters using an Artificial Neural Network with a QuadTree-based Feature Extractor}",
      journal = {\aj},
         year = 2024,
        month = jan,
       volume = {167},
       number = {1},
          eid = {12},
        pages = {12},
          doi = {10.3847/1538-3881/ad07e5},
archivePrefix = {arXiv},
       eprint = {2311.03009},
 primaryClass = {astro-ph.GA},
       adsurl = {https://ui.adsabs.harvard.edu/abs/2024AJ....167...12C}
}

@ARTICLE{Nguyen2022,
       author = {{Nguyen}, C.~T. and {Costa}, G. and {Girardi}, L. and {Volpato}, G. and {Bressan}, A. and {Chen}, Y. and {Marigo}, P. and {Fu}, X. and {Goudfrooij}, P.},
        title = "{PARSEC V2.0: Stellar tracks and isochrones of low- and intermediate-mass stars with rotation}",
      journal = {\aap},
         year = 2022,
        month = sep,
       volume = {665},
          eid = {A126},
        pages = {A126},
          doi = {10.1051/0004-6361/202244166},
archivePrefix = {arXiv},
       eprint = {2207.08642},
 primaryClass = {astro-ph.SR},
       adsurl = {https://ui.adsabs.harvard.edu/abs/2022A&A...665A.126N}
}

@ARTICLE{Riello2021,
       author = {{Riello}, M. and {De Angeli}, F. and {Evans}, D.~W. and {Montegriffo}, P. and {Carrasco}, J.~M. and {Busso}, G. and {Palaversa}, L. and {Burgess}, P.~W. and {Diener}, C. and {Davidson}, M. and {Rowell}, N. and {Fabricius}, C. and {Jordi}, C. and {Bellazzini}, M. and {Pancino}, E. and {Harrison}, D.~L. and {Cacciari}, C. and {van Leeuwen}, F. and {Hambly}, N.~C. and {Hodgkin}, S.~T. and {Osborne}, P.~J. and {Altavilla}, G. and {Barstow}, M.~A. and {Brown}, A.~G.~A. and {Castellani}, M. and {Cowell}, S. and {De Luise}, F. and {Gilmore}, G. and {Giuffrida}, G. and {Hidalgo}, S. and {Holland}, G. and {Marinoni}, S. and {Pagani}, C. and {Piersimoni}, A.~M. and {Pulone}, L. and {Ragaini}, S. and {Rainer}, M. and {Richards}, P.~J. and {Sanna}, N. and {Walton}, N.~A. and {Weiler}, M. and {Yoldas}, A.},
        title = "{Gaia Early Data Release 3. Photometric content and validation}",
      journal = {\aap},
         year = 2021,
        month = may,
       volume = {649},
          eid = {A3},
        pages = {A3},
          doi = {10.1051/0004-6361/202039587},
archivePrefix = {arXiv},
       eprint = {2012.01916},
 primaryClass = {astro-ph.IM},
       adsurl = {https://ui.adsabs.harvard.edu/abs/2021A&A...649A...3R}
}

@ARTICLE{Chen2019c,
       author = {{Chen}, Yang and {Girardi}, L{\'e}o and {Fu}, Xiaoting and {Bressan}, Alessandro and {Aringer}, Bernhard and {Dal Tio}, Piero and {Pastorelli}, Giada and {Marigo}, Paola and {Costa}, Guglielmo and {Zhang}, Xing},
        title = "{YBC: a stellar bolometric corrections database with variable extinction coefficients. Application to PARSEC isochrones}",
      journal = {\aap},
         year = 2019,
        month = dec,
       volume = {632},
          eid = {A105},
        pages = {A105},
          doi = {10.1051/0004-6361/201936612},
archivePrefix = {arXiv},
       eprint = {1910.09037},
 primaryClass = {astro-ph.SR},
       adsurl = {https://ui.adsabs.harvard.edu/abs/2019A&A...632A.105C}
}

@ARTICLE{Li2020a,
       author = {{Li}, Lu and {Shao}, Zhengyi and {Li}, Zhao-Zhou and {Yu}, Jincheng and {Zhong}, Jing and {Chen}, Li},
        title = "{Modeling Unresolved Binaries of Open Clusters in the Color-Magnitude Diagram. I. Method and Application of NGC 3532}",
      journal = {\apj},
         year = 2020,
        month = sep,
       volume = {901},
       number = {1},
          eid = {49},
        pages = {49},
          doi = {10.3847/1538-4357/abaef3},
archivePrefix = {arXiv},
       eprint = {2008.04684},
 primaryClass = {astro-ph.GA},
       adsurl = {https://ui.adsabs.harvard.edu/abs/2020ApJ...901...49L}
}

@ARTICLE{Li2025,
      title={The MiMO Catalog: Physical Parameters and Stellar Mass Functions of 1,232 Open Clusters from Gaia DR3}, 
      author={Lu Li and Zhengyi Shao and Zhaozhou Li and Xiaoting Fu},
      year={2025},
      eprint={2510.23374},
      archivePrefix={arXiv},
      primaryClass={astro-ph.GA},
      url={https://arxiv.org/abs/2510.23374}, 
}
\bibliographystyle{aasjournal}

\end{document}